\documentclass[%
superscriptaddress,
twocolumn,
showpacs,preprintnumbers,
 amsmath,amssymb,
 aps,
prb,
]{revtex4-2}

\usepackage{graphicx}
\usepackage{dcolumn}
\usepackage{bm}
\usepackage[colorlinks,
linkcolor=blue, urlcolor=blue, anchorcolor=blue, citecolor=blue]{hyperref}
\usepackage{float}
\usepackage{multirow}
\usepackage{xcolor}
\usepackage[normalem]{ulem}

\begin{document}


\title{Kitaev interaction and possible spin liquid state in CoI$_2$ and Co$_{2/3}$Mg$_{1/3}$I$_2$}

\author{Yaozhenghang Ma}
\thanks{Y. M. and K. Y. contributed equally to this work.}
 \affiliation{Laboratory for Computational Physical Sciences (MOE),
 State Key Laboratory of Surface Physics, and Department of Physics,
 Fudan University, Shanghai 200433, China}

\author{Ke Yang}
\thanks{Y. M. and K. Y. contributed equally to this work.}
\affiliation{College of Science, University of Shanghai for Science and Technology, Shanghai 200093, China}
 \affiliation{Laboratory for Computational Physical Sciences (MOE),
 State Key Laboratory of Surface Physics, and Department of Physics,
 Fudan University, Shanghai 200433, China}

\author{Yuxuan Zhou}
 \affiliation{Laboratory for Computational Physical Sciences (MOE),
 State Key Laboratory of Surface Physics, and Department of Physics,
 Fudan University, Shanghai 200433, China}

 \author{Hua Wu}
 \email{Corresponding author. wuh@fudan.edu.cn}
 \affiliation{Laboratory for Computational Physical Sciences (MOE),
  State Key Laboratory of Surface Physics, and Department of Physics,
  Fudan University, Shanghai 200433, China}
 \affiliation{Shanghai Qi Zhi Institute, Shanghai 200232, China}
 \affiliation{Hefei National Laboratory, Hefei 230088, China}

\date{\today}

\begin{abstract}

Kitaev materials are of great interest due to their potential in realizing quantum spin liquid (QSL) states and applications in topological quantum computing. In the pursuit of realizing Kitaev QSL, a Mott insulator with strong bond-dependent frustration and weak geometric frustration is highly desirable. Here we explore Kitaev physics in the van der Waals triangular antiferromagnet (AF) CoI$_2$, through the spin-orbital states and Wannier function analyses, exact diagonalization and density matrix renormalization group study of the electronic structure and magnetic properties. We find that the high-spin Co$^{2+}$ ion is in the $J_\mathrm{eff}=1/2$ state because of strong spin-orbit coupling, and the weak trigonal elongation and crystal field contribute to the observed weak in-plane magnetic anisotropy. The strong $t_{2g}$-$e_g$ hopping via the strong Co 3$d$-I 5$p$ hybridization gives rise to a strong Kitaev interaction ($K_1$) at the first nearest neighbors (1NN), and the long Co-Co distance and the weak $t_{2g}$-$t_{2g}$ hoppings determine a weak Heisenberg interaction $J_1$. The resultant $|K_1/J_1|$ = 6.63 confirms a strong bond-dependent frustration, while the geometric frustration due to the 3NN Heisenberg interaction $J_3$ gets involved, and they all together result in the experimental helical AF order in CoI$_2$. We then propose to suppress the $J_3$ using a partial Mg substitution for Co, and indeed we find that Co$_{2/3}$Mg$_{1/3}$I$_2$ has the much reduced geometric frustration but hosts the robust bond-dependent frustration, and thus it would be a promising Kitaev material being so far closest to the QSL state.

\end{abstract}

\maketitle


Quantum spin liquid (QSL) is one of the most exotic phases in condensed matter physics and has attracted significant interest since it was first conceptualized by Anderson in the 1970s \cite{Anderson1973, Fazekas1974}. Considerable efforts have been made to realize quantum spin liquids utilizing geometric frustration on triangular, Kagome, and pyrochlore lattices \cite{Balents2010}. Recently, the Kitaev model, which incorporates bond-dependent frustration, has garnered considerable attention due to its Kitaev QSL ground state and Majorana fermion excitations \cite{Kitaev2006}. Jackeli and Khaliullin proposed a method to realize the Kitaev model in $d^5$ transition metal compounds through the assistance of strong spin-orbit coupling (SOC) \cite{Jackeli2009, Takagi2019}. Significant efforts have been devoted to realizing the Kitaev QSL in $4d^5$ and $5d^5$ systems, such as $\alpha$-RuCl$_3$ \cite{Plumb2014, Sandilands2015, Kim2015, Banerjee2016} and Na$_2$IrO$_3$ \cite{Chaloupka2010, Liu2011, Ye2012, Chaloupka2013, Hwan2015}. However, discovering a quantum spin liquid material remains challenging due to the non-negligible Heisenberg interaction. To reduce Heisenberg  interaction, Liu \textit{et al} \cite{Liu2018} and Sano \textit{et al} \cite{Sano2018} proposed that in $3d^7$ systems the contributions to Heisenberg interaction from $t_{2g}$-$e_g$ channels and $e_{g}$-$e_{g}$ channels cancel each other, while the $t_{2g}$-$e_g$ channels contribute to a dominant ferromagnetic (FM) Kitaev interaction. Co$^{2+}$-based $3d^7$ materials, such as BaCo$_2$(AsO$_4$)$_2$ \cite{Zhong2020, Chen2021, Das2021, Maksimov2022, Winter2022, Halloran2023,  Liu2023}, Na$_3$Co$_2$SbO$_6$ \cite{Liu2020, Songvilay2020, Kim2021, Sanders2022, Li2022, Vavilova2023} and Na$_2$Co$_2$TeO$_6$ \cite{Songvilay2020, Lin2021, Kim2021, Sanders2022, Yao2022}, have been extensively studied in the quest for a dominant Kitaev interaction. However, the study by Liu and Kee reveals that the contributions from $t_{2g}$-$t_{2g}$ channels are non-negligible, which boost the otherwise canceled Heisenberg interaction \cite{Liu2023}.

\begin{figure*}[t]
  \includegraphics[width=\textwidth]{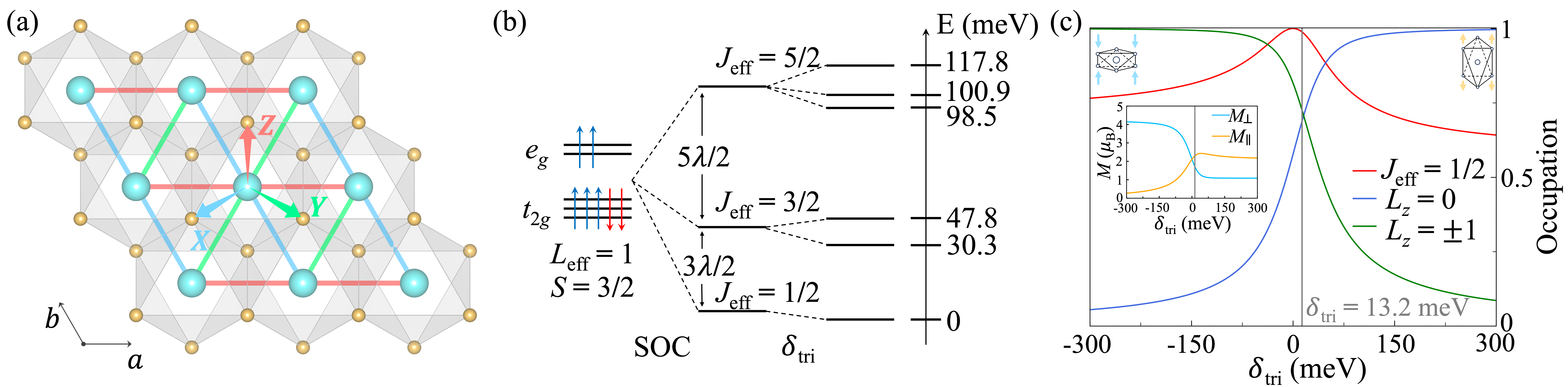}
  \centering
  \caption{(a) The $c$-axis view of crystal structure and local basis of CoI$_2$. The cyan and yellow balls correspond to Co and I atoms. The local X, Y and Z-axis and the corresponding Kitaev bonds of each axis are indicated by blue, green and red color. (b) Energy splitting of $d^7$ high spin state under SOC and trigonal crystal field. The excited energy levels are predicted by ED calculation. (c) The occupation number of ground state under different trigonal crystal field.  The inset shows out-of-plane (blue curve) and in-plane (yellow curve) magnetic moments under different trigonal crystal field. The gray line denotes trigonal crystal field strength $\delta$ of CoI$_2$ in our calculation.
  }
\label{structure}
\end{figure*}

Very recently, Kim \textit{et al} reported bond-dependent anisotropy in the van der Waals antiferromagnet CoI$_2$ \cite{Kim2023}. CoI$_2$ contains a triangular network of Co ions with the edge-sharing CoI$_6$ octahedra, as shown in Fig.\ref{structure}(a). It exhibits a helical antiferromagnetic (AF) order below the N\'eel temperature $T_\mathrm{N}\approx9\ \mathrm{K}$ \cite{Kuindersma1981, Kurumaji2013, Kim2023}. The $3d^7$ Co$^{2+}$ ion could adopt a $S=3/2$ high spin state, and the hole on $t_{2g}$ states would contribute to $L=1$ effective angular momentum. Then the SOC would mix the multiplets into $J_{\mathrm{eff}}$ states which are desirable for the Kitaev physics. However, the high spin state may be destabilized, if the energy gain from Hund's coupling is insufficient to counteract the energy cost in octahedral crystal field. In addition, the $J_{\mathrm{eff}}$ states from SOC would be disfavored, if the trigonal crystal field is strong enough to destroy the effective $L=1$ states. Therefore, it is of great interest to investigate whether the Co$^{2+}$ ion is in the $J_{\mathrm{eff}}=1/2$ Kramers doublet and then produces significant Kitaev interactions. Note also that compared with those previously studied Co$^{2+}$ materials, CoI$_2$ has a more open structure and larger Co-Co distance which would weaken the Heisenberg interactions. Moreover, the heavy ligand iodine atoms help to enhance the Kitaev interactions \cite{Xu2018, Xu2020}. Thus possibly, CoI$_2$ has dominant Kitaev interactions to achieve QSL.

Based on the above picture, we study the electronic structure and magnetic properties of CoI$_2$, using first-principles calculation, Wannier function analysis, exact diagonalization (ED) and density matrix renormalization group (DMRG). Our work confirms that the SOC effect is stronger than trigonal crystal field, and the electronic ground state is $J_{\mathrm{eff}}=1/2$ state with weak in-plane magnetic anisotropy. Our study shows that the $t_{2g}$-$e_g$ channels contribute to a strong Kitaev interaction and the Heisenberg interaction from $t_{2g}$-$t_{2g}$ channels is effectively suppressed. However, with involvement of the considerably strong antiferromagnetic (AFM) interaction $J_3$ between third nearest neighbors, the geometric frustration results in the helical magnetic state as experimentally observed. Therefore, we propose a partial Mg substitution for Co to block the $J_3$ and thus suppress the geometric frustration, and then indeed we find that  Co$_{2/3}$Mg$_{1/3}$I$_2$ has the overwhelming Kitaev interaction and most closely approaches the Kitaev QSL among several candidates in literature.

To study the electronic structure, we employ a multi-orbital Hubbard model incorporating the five $d$ orbitals:

\begin{equation}
\label{eq:one-site}
	H_{\mathrm{one}} = H_{\mathrm{CF}} + H_{U} + H_{\mathrm{SOC}}
\end{equation}
where $H_{\mathrm{CF}}$, $H_{U}$, and $H_{\mathrm{SOC}}$ represent crystal field splitting, Coulomb repulsion and spin-orbit coupling, respectively. The crystal field term for the Co$^{2+}$ ion can be expressed in the basis $\mathbf{c} = (d_{3z^2 - r^2}, d_{x^2 - y^2}, d_{xz}, d_{yz}, d_{xy})$ as
\begin{equation}
 	H_{\mathrm{CF}}^{\mathrm{Co}}=\begin{pmatrix}\Delta&0&0&0&0\\0&\Delta&0&0&0\\0&0&0&\delta&\delta\\0&0&\delta&0&\delta\\0&0&\delta&\delta&0\end{pmatrix},
\end{equation}
where $\Delta$ represents the energy difference between the $e_g$ and $t_{2g}$ orbitals, and $\delta_{\rm {tri}}$ = $3\delta$ denotes the splitting between the $a_{1g}$ and $e^\pi_{g}$ orbitals as derived from the trigonal distortion in the [111] direction of octahedron. The crystal field details are provided in Supplementary Material (SM) I \cite{SM}. The Coulomb repulsion is modeled using isotropic Kanamori form \cite{Georges2013}. The SOC term is given by $H_{\mathrm{SOC}} = \lambda \mathbf{L}\cdot \mathbf{S}$, where $\lambda = 26.6$ meV is the atomic SOC strength. We calculated electronic structure and magnetic moment by ED method, as shown in SM II \cite{SM}.

The magnitude of the $t_{2g}$-$e_g$ octahedral crystal field splitting ($\Delta$) relative to $J_{\rm H}$ determines the spin state of the Co$^{2+}$ ion. For estimation purposes: the $S = 3/2$ state ($3d^{5\uparrow}t_{2g}^{2\downarrow}$) has a Hund's coupling energy of $-11J_{\rm H}$ plus $2\Delta_{\rm cf}$ (the crystal field excitation energy of two electrons in the $e_g$ orbitals), while the $S = 1/2$ state ($t_{2g}^{3\uparrow 3\downarrow} e_g^{1\uparrow}$) has a total stabilization energy of $-9J_{\rm H}$ plus $\Delta_\mathrm{cf}$. Thus, the critical condition for a high-spin to low-spin transition is $\Delta_{\rm cf} > 2J_{\rm H} = 1.8$ eV. Our Wannier analysis yields a $t_{2g}$-$e_g$ crystal field splitting of $\Delta = 1.02$ eV, and much smaller than the critical value of 1.8 eV. Therefore, CoI$_2$ is well stabilized in the high-spin $S = 3/2$ ($3d^{5\uparrow}t_{2g}^{2\downarrow}$) ground state.

In the ($3d^{5\uparrow}t_{2g}^{2\downarrow}$) ground state, the $t_{2g}$ orbitals are not fully occupied, resulting in an unquenched orbital angular momentum with $L_{\mathrm{eff}} = 1$. Consequently, this spin-orbital multiplet comprises 12-degenerate states characterized by ($L_{\mathrm{eff}} = 1$, $S = 3/2$).
When SOC is introduced, these 12-degenerate states further split into $J_{\mathrm{eff}} = 1/2$ , $J_{\mathrm{eff}} = 3/2$ and $J_{\mathrm{eff}} = 5/2$ states, as shown in Fig.\ref{structure}(b).
Moreover, under the influence of a trigonal crystal field, the ground state may also be in $L_z = 0$ or $L_z = \pm 1$ state, as shown in Fig.\ref{structure}(c).
Thus, the competition between spin-orbit coupling and trigonal crystal field splitting ultimately determines the intriguing spin-orbital states.

For the Co$^{2+}$ high-spin (HS) state studied here, the spin-orbit coupling strength is typically fixed. Therefore, we vary the trigonal crystal field $\delta_{\rm {tri}}$ and use ED to determine how the ground state evolves. To visualize the changes in the ground state, we project the ED-obtained ground state wavefunctions onto the possible basis states of $J_{\mathrm{eff}} = 1/2$, $L_z = \pm 1$, and $L_z = 0$.
As shown in Fig.~\ref{structure}(c), it is evident that when $\delta_{\rm {tri}}$ is less than $-150$~meV, corresponding to a compressed trigonal distortion, the $a_{1g}$ orbital energy decreases, and the orbital ground state is $L_z = \pm 1$. Conversely, when $\delta_{\rm {tri}}$ exceeds $150$~meV, corresponding to an elongated trigonal distortion, the $a_{1g}$ orbital energy increases, and the orbital ground state becomes $L_z = 0$. Notably, only within the narrow range of $-38$~meV $< \delta_{\rm {tri}} < 48$~meV does the system exhibit a well-defined $J_{\mathrm{eff}} = 1/2$ ground state, where the two boundary values are determined by the two crossing points of the  three curves of $J_{\mathrm{eff}} = 1/2$, $L_z = \pm 1$, and $L_z = 0$ in Fig. \ref{structure}(c).
To double check accuracy of the obtained ground state, we calculate the expectation value of the total magnetic moment operator, $L + 2S$, from the ground state wavefunction. The results indicate that, when the trigonal field is zero (i.e., under a cubic crystal field), the calculated in-plane and out-of-plane magnetic moments have the equal value of $\mathrm{2.11}\ \mu_\mathrm{B}$, and the associated L\'and $g$-factor is 4.22. This is in good agreement with the experimental value $g=4.28$ for Co$^{2+}$ in the cubic MgO lattice \cite{Low1962, Abragam2012}.

Then, using maximally localized Wannier functions (MLWFs) generated with Wannier90 \cite{wannier90}, we obtained the trigonal crystal field splitting of \( \delta_{\mathrm{tri}} \) = 13.2 meV for the CoI$_2$. By projecting the ground-state wavefunction onto the \( J_{\mathrm{eff}} = 1/2 \) state, we found that the weight reaches as high as 0.98. Note that the small deviation from the pure \( J_{\mathrm{eff}} \) = 1/2 state results from the trigonal distortion.
We also calculated the $g$-factor for the weak trigonal distortion in CoI$_2$: the in-plane $g$-factor is \( g_{||}=4.62 \), while the out-of-plane $g$-factor is \( g_{\perp}=3.43 \). The anisotropic $g$-factor indicates that the \( ab \)-plane is the easy plane for CoI$_2$, which is in agreement with experiments \cite{Kuindersma1981,Kurumaji2013}.
Moreover, due to this weak trigonal distortion, the \( J_{\mathrm{eff}} = 3/2 \) and \( J_{\mathrm{eff}} = 5/2 \) excited states further split into five doublets. Our ED calculation gives the excitation energies as \{30.3, 47.8, 98.5, 100.9, 117.8\} meV. The first excitation energy of 30.3 meV is consistent with inelastic neutron scattering data, which reports a value of 35.4 meV \cite{Kim2023}. The rest higher-energy excitation states call for experimental verification.

We determined that the electronic structure of CoI$_2$ is described by $J_{\mathrm{eff}}=1/2$ picture, and thus the CoI$_2$ may host the Kitaev interaction. In our calculation, we found the electronic excitation from $J_{\mathrm{eff}}=1/2$ to $J_{\mathrm{eff}}=3/2$ requires at least 30.3 meV, which is above the room temperature. Thus, the energy spectrum safely restricts to $J_{\mathrm{eff}}=1/2$ states in our work. And magnetic structure of CoI$_2$ can be studied within $J_{\mathrm{eff}}=1/2$ Kramers doublet framework using a low-energy $S=1/2$ spin Hamiltonian.

To study the magnetic exchange parameters, we employ a multi-orbital Hubbard model including two Co$^{2+}$ sites:
\begin{equation}
\label{eq:Hubbard}
	H = H_{\mathrm{one}} + H_{t},
\end{equation}
where $H_{\mathrm{one}}$ and $H_{t}$ represent one-site Hamiltonian term of each Co$^{2+}$ and hopping terms between two Co$^{2+}$, respectively. The $Z$ bond hopping matrix between sites Co$_i$ and Co$_j$ can be expressed in the basis $\mathbf{c} = (d_{3z^2 - r^2}, d_{x^2 - y^2}, d_{xz}, d_{yz}, d_{xy})$ as
\begin{equation}
    H_t^{(ij)}=\begin{pmatrix}t_5&0&0&0&t_6\\0&t_4&0&0&0\\0&0&t_1&t_2&0\\0&0&t_2&t_1&0\\t_6&0&0&0&t_3\end{pmatrix},
\end{equation}
where $t_{1-6}$ are the hopping terms allowed by $C_{2v}$ symmetry \cite{Liu2023}. These hopping terms and all other non-zero terms due to the lowered $C_s$ symmetry are obtained through MLWFs, as shown in SM III \cite{SM}. The magnetic properties are studied within a low-energy spin Hamiltonian, in which the $J_{\mathrm{eff}}=1/2$ Kramers doublets are mapped into $S=1/2$ pseudospin. For the triangular and hexagonal lattice, the general spin Hamiltonian is given by
\begin{equation}
\label{HK_Hamiltonian}
\begin{aligned}
     H = \sum_{ij} &J \mathbf{S}_i \cdot \mathbf{S}_j + K S_i^\gamma S_j^\gamma + \Gamma \left(S_i^\alpha S_j^\beta + S_i^\beta S_j^\alpha \right) \\
     &+ \Gamma^{\prime} \left(S_i^\gamma S_j^\alpha + S_i^\gamma S_j^\beta + S_i^\alpha S_j^\gamma + S_i^\beta S_j^\gamma \right)
\end{aligned}
\end{equation}
where \{$\alpha, \beta, \gamma$\} = \{$x, y, z$\}, \{$y, z, x$\}, and \{$z, x, y$\} for the $Z$, $X$, and $Y$ bonds, respectively. The spin Hamiltonian is obtained by solving the multi-orbital Hubbard model, and the details of calculation are shown in SM IV \cite{SM}.

\begin{figure}[t]
  \includegraphics[width=8cm]{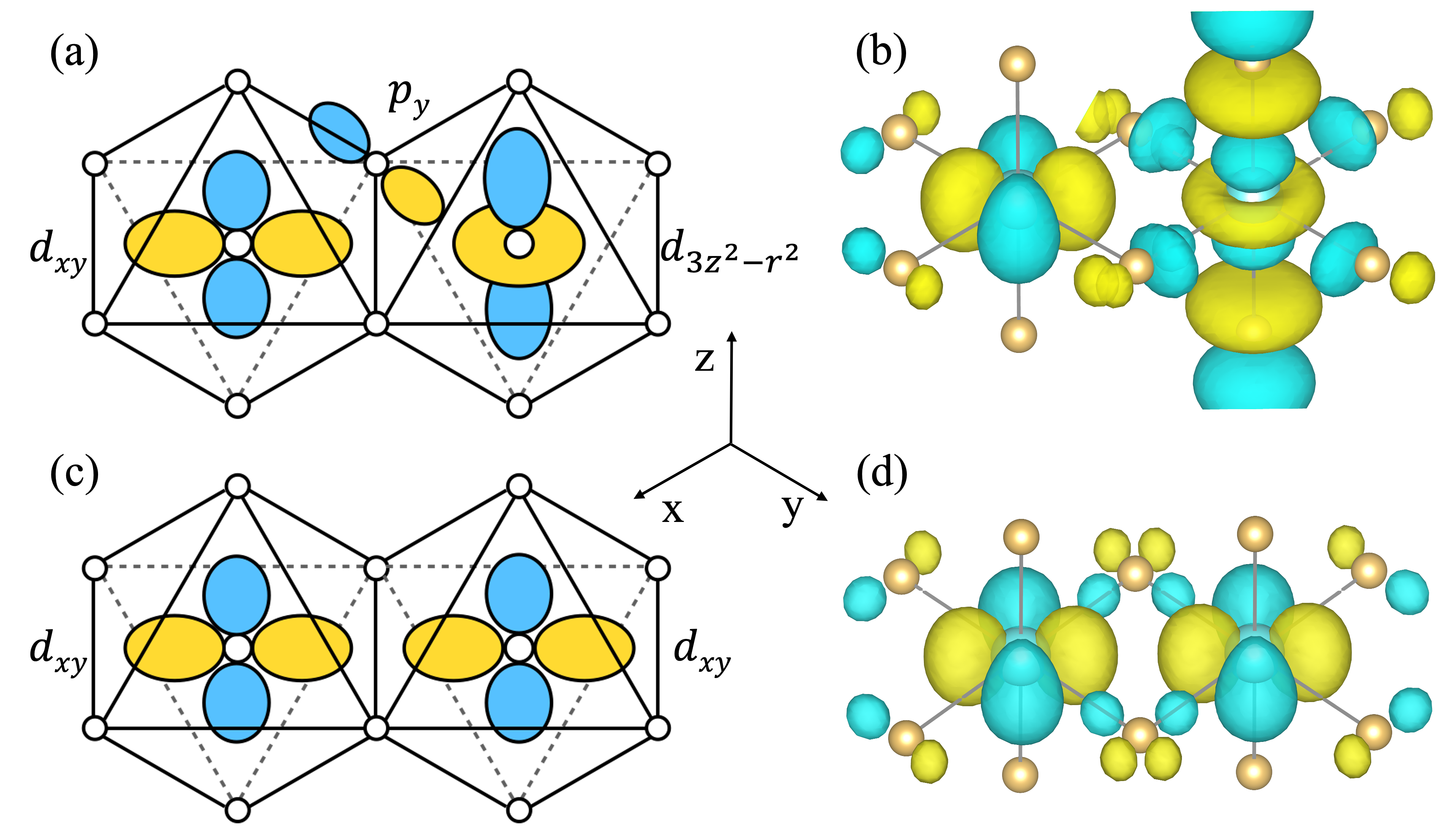}
  \centering
  \caption{The schematic diagram of (a) the dominant $t_{2g}$-$e_g$ hopping: indirect hopping between $d_{xy}$ and $d_{3z^2-r^2}$ orbitals and (c) the dominant $t_{2g}$-$t_{2g}$ hopping: direct hopping between $d_{xy}$ orbitals. The corresponding Wannier functions are plotted in (b) and (d).
  }
\label{two_hopping}
\end{figure}

For CoI$_2$, the interaction between first nearest neighbors is the most important term, and the strongest hopping channel is $t_6=177.1\;\mathrm{meV}$, as shown in Eq. S17 of SM \cite{SM}. As shown in Fig. \ref{two_hopping}(a)-(b), the strongest $t_6$ term is indirect hopping term between $t_{2g}$ and $e_g$ orbitals, which makes difference with other Co$^{2+}$ materials. And unlike others, in CoI$_2$ the $t_{2g}$-$t_{2g}$ hopping channel, as illustrated in Fig. \ref{two_hopping}(c)-(d), is relative weak with value $t_3=-44.1\;\mathrm{meV}$ in Eq. S17 of SM \cite{SM}. The Co-Co distance $3.87\;\mathrm{\AA}$ in CoI$_2$ is much longer than $2.89\;\mathrm{\AA}$ in BaCo$_2$(AsO$_4$)$_2$ \cite{Halloran2023} and $2.80\;\mathrm{\AA}$ in BaCo$_2$(PO$_4$)$_2$ \cite{Nair2018}. The longer bond suppresses the direct hopping channels, leading to relative enhancement of indirect hopping channels. Additionally, CoI$_2$ has the heavier ligand iodine, and the strong interaction between Co$^{2+}$ $3d$ orbitals and I$^-$ $5p$ orbitals enhances the $p$-orbital-mediated hopping process $t_6$. As shown in Table S1 of SM\cite{SM}, the dominant $t_6$ channel contributes to a strong FM Kitaev interaction $K=-4.73\; \mathrm{meV}$ and a weak AFM Heisenberg interaction $J=0.81\; \mathrm{meV}$, leading to the dominant Kitaev interaction in CoI$_2$.

For triangular \cite{Balents2010} and hexagonal lattices \cite{Halloran2023}, the geometric frustration associated with second and third nearest neighbors' interaction attracts considerable attention, and it turns out to be important as seen below. The complex competition between geometric frustration and bond dependent frustration determines the magnetic ground state. As shown in Table S1 and Fig. S9 of SM \cite{SM}, the geometric frustration arising from second nearest neighbors is negligible compared to that originating from third nearest neighbors, where the dominant channel $t_2=-39.1\;\mathrm{meV}$ associated with second nearest neighbors only contributes $J=-0.24\;\mathrm{meV}$ while the principle channel $t_4=92.0\;\mathrm{meV}$ involving in third nearest neighbors yields a significantly larger contribution with $J=1.9\;\mathrm{meV}$. Thus the AFM Heisenberg interaction between third nearest neighbors is the the primary contributor to geometric frustration.

\begin{table}[t]
\centering
\setlength{\tabcolsep}{1.2mm}
\caption{First, second and third nearest neighbor magnetic interactions in meV for CoI$_2$ (Co$_{2/3}$Mg$_{1/3}$I$_2$).}
\label{hopping_table}
\begin{tabular}{lcccc}
  \hline \hline
    &  $J$            &  $K$             & $\Gamma$       &  $\Gamma^\prime$ \\ \hline
1NN & 0.63 (0.89) & -4.17 (-3.15)  & -0.04 (-0.07)  & 0.06  (0.27) \\
2NN & -0.15 (-0.21) & -0.56 (-0.32)  & 0.40 (0.04)    & -0.14 (0.02) \\
3NN & 2.16 (0.53)   & -0.06 (0.01)   & 0.38 (0.01)    & 0.30 (0.06) \\ \hline \hline
\end{tabular}
\end{table}

Then we solved the multi-orbital Hubbard model containing all hopping channels and obtained the exchange parameters in the spin Hamiltonian, shown in Table \ref{hopping_table}. As proposed above, we found a strong FM Kitaev interaction $K_1= -4.17\; \mathrm{meV}$ and a relative weak AFM Heisenberg interaction $J_1= 0.63\; \mathrm{meV}$. The bond-dependent frustration in CoI$_2$ is remarkable, with the $|K_1|/|J_1|$ ratio equal to 6.63, while the ratio in BaCo$_2$(AsO$_4$)$_2$ is only 0.06 \cite{Liu2023}. Meanwhile, the strong AFM Heisenberg interaction $J_3 = 2.16 \; \mathrm{meV}$ induces strong geometric frustration.

\begin{figure}[t]
  \includegraphics[width=8.5cm]{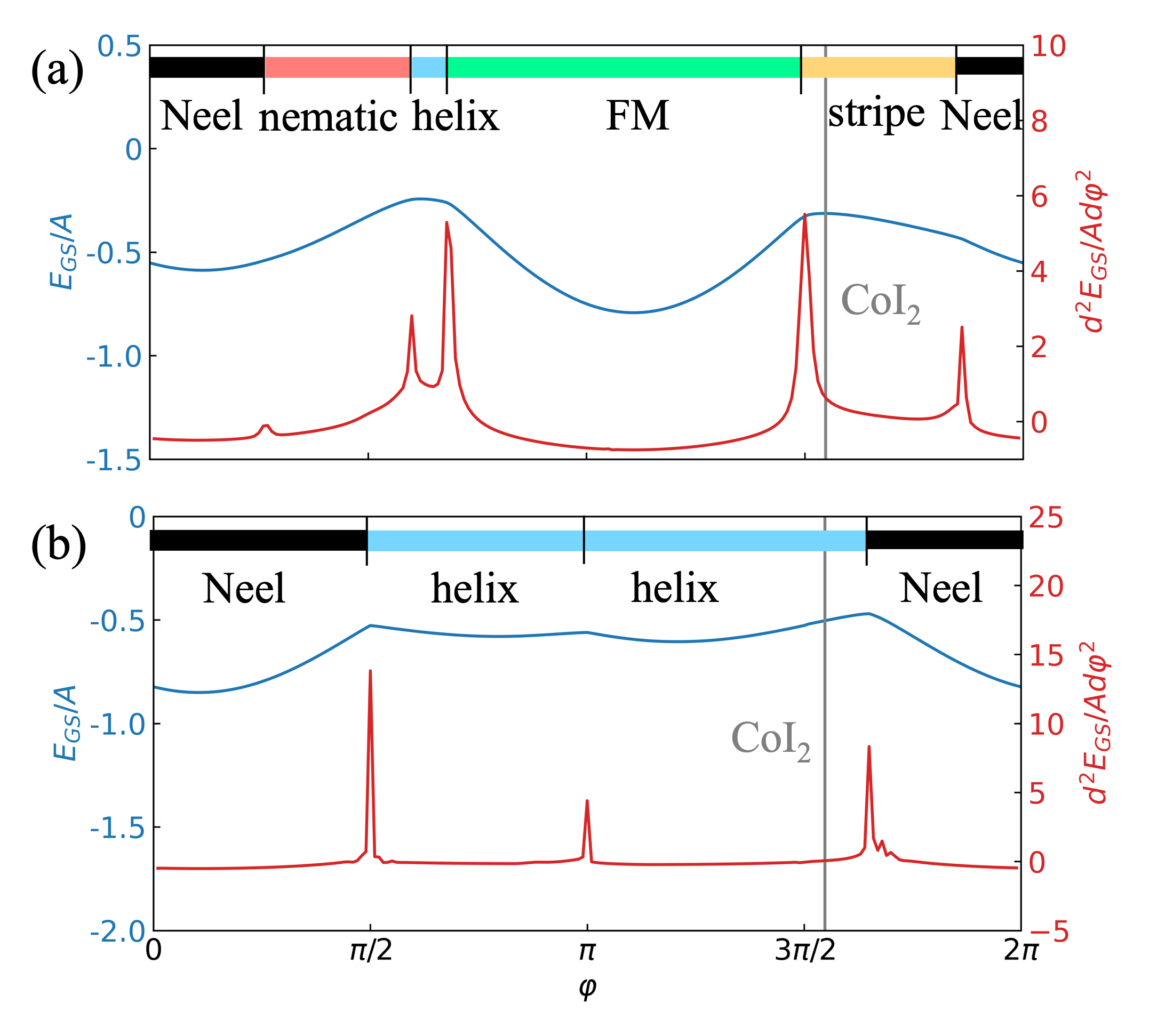}
  \centering
  \caption{The DMRG results and phase diagrams of Kitaev-Heisenberg model on triangular lattice (a) without $J_3$ and (b) with $J_3$. The ground state energy $E_\mathrm{GS}/A$ per site and its second derivative $d^2E_\mathrm{GS}/Ad\varphi^2$ are indicated by blue and red line respectively.
  }
\label{phase}
\end{figure}

The existence of bond-dependent and geometric frustration suggests that the magnetic ground state is determined by their complex competition. Here, we used DMRG method to calculate the phase diagram with and without $J_3$ on a $6\times6$ triangular lattice. The energy values in our DMRG calculation without $J_3$ are consistent with those derived from ED calculation on $6\times4$ triangular lattice in Fig. 10 of reference \cite{Becker2015}, which corroborates the accuracy of our DMRG calculation. To simplify the model Hamiltonian, we included the three dominant interactions in phase diagram, Heisenberg interaction $J_1$, Kitaev interaction $K_1$ and Heisenberg interaction $J_3$. As shown in Fig. \ref{phase}, in our calculations, we set $J_1 = A \cos \varphi$ and $K_1 = A \sin \varphi$ using the angle parameter $\varphi \in [0, 2\pi]$, with $A=\sqrt{K_1^2+J_1^2}$.

For the parameter space without $J_3$, the phase diagram is divided into five regions, as shown in Fig. \ref{phase}(a). For a pure Heisenberg model, the ground state is either a FM state or a $120^{\circ}$ N\'eel AFM state. When the Kitaev interaction is included, the bond-dependent frustration induces three new phases: helical AFM, nematic, and stripe AFM. In total, the ground state turns out to be the stripe AFM but not the experimental helical AFM. However, in the parameter space with $J_3$, the bond-dependent frustration is suppressed by geometric frustration, leaving only the $120^{\circ}$ Neel AFM state and the helical AFM state, as shown in Fig. \ref{phase}(b). Now the ground state becomes the helical AFM state in agreement with the experiment ~\cite{Kim2023}. Therefore, both the FM Kitaev interaction $K_1$ and the AFM Heisenberg interaction $J_3$ play a major role in deterring the experimental helical AFM state of CoI$_2$~\cite{Kim2023}.

\begin{figure}[t]
  \includegraphics[width=8cm]{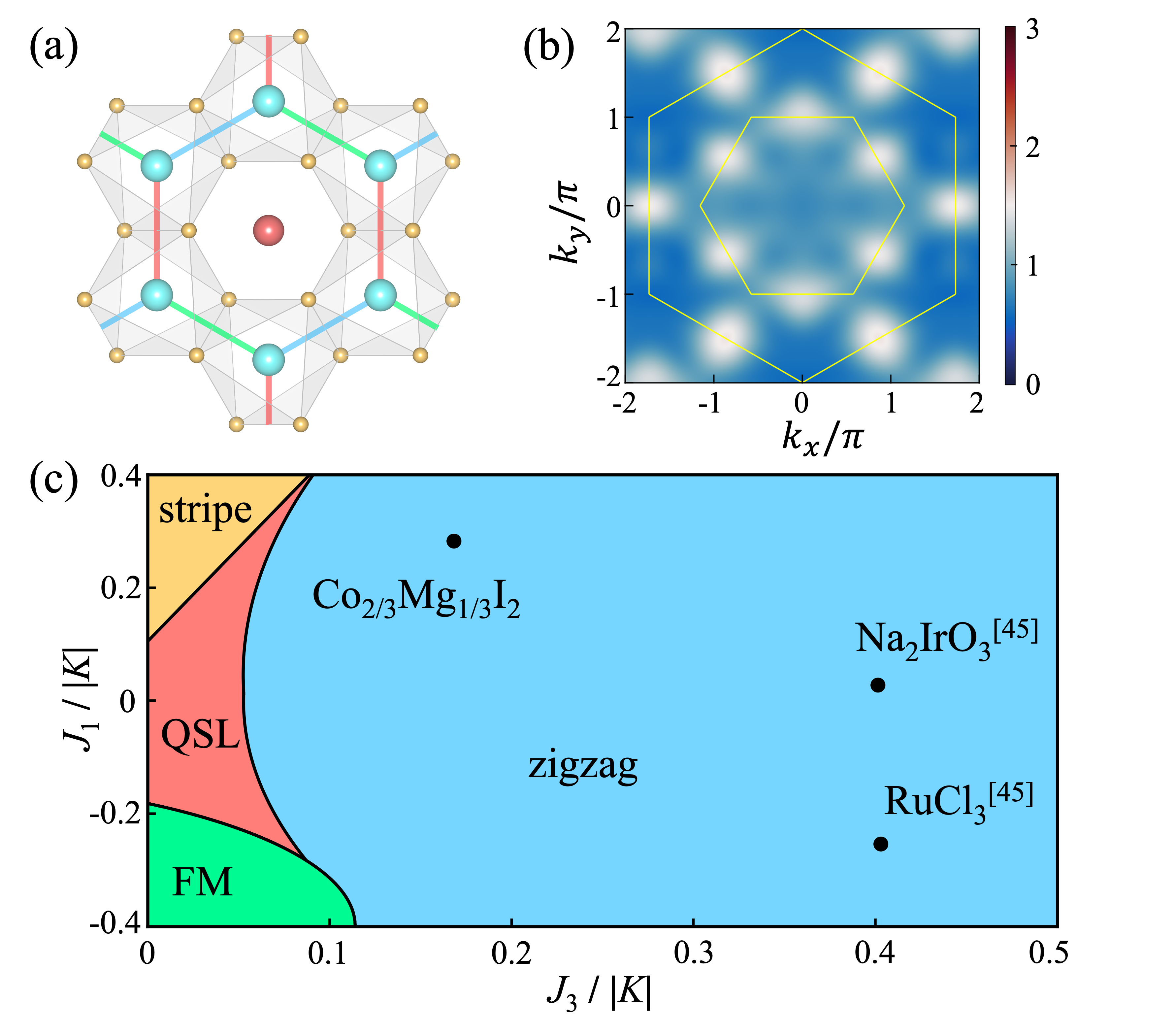}
  \centering
  \caption{(a) The crystal structure of Co$_{2/3}$Mg$_{1/3}$I$_2$. The cyan, red and yellow balls represent Co, Mg and I atoms respectively. (b) Magnetic structure factor $S(q)$ of Co$_{2/3}$Mg$_{1/3}$I$_2$ with $30\%$ $J_3$. The colormap is limited to the range of 0 to the maximum value $S(q)_\mathrm{max}$ of zigzag and stripe order as shown in Fig. S11 of SM \cite{SM}. (c) Phase diagram for FM Kitaev interaction $K<0$ by ED. The exchange parameters of Na$_2$IrO$_3$ and RuCl$_3$ are from Ref. \cite{Winter2016}.
  }
\label{CoMgI}
\end{figure}

The bond-dependent frustration promotes a Kitaev QSL state when the Kitaev interaction dominates, while the geometric frustration arising from longer-range Heisenberg terms tends to stabilize non-QSL orders such as the helical AFM phase. To explore the bond-dependent QSL, we propose Co$_{2/3}$Mg$_{1/3}$I$_2$ (with a partial Mg substitution for Co, see Fig. \ref{CoMgI}(a)) to block the $J_3$ interaction via intermediate Mg ion. Thus, the suppression of the $J_3$-driven geometric frustration would unmask the bond-dependent frustration. Note that the induced trigonal distortion remains small, as Mg$^{2+}$ has similar ionic size of $0.72\; \mathrm{\AA}$ to that of $0.745\; \mathrm{\AA}$ for Co$^{2+}$ \cite{Shannon1976}.

For Co$_{2/3}$Mg$_{1/3}$I$_2$, we performed another set of DFT, MLWFs and ED calculations. Here the calculated octahedral $t_{2g}$-$e_g$ splitting of 0.99 eV (after structural relaxation) is almost the same as that of 1.02 eV in CoI$_2$, and the trigonal crystal field splitting remains tiny (here only 4.5 meV), thus resulting in the robust $J_{\mathrm{eff}}=1/2$ ground state in Co$_{2/3}$Mg$_{1/3}$I$_2$. The interaction between first nearest neighbors remains dominated by the $t_6$ term, while the third nearest neighbors' contributions are notably suppressed, as seen in Table \ref{hopping_table}. Now the $|K_1|/|J_3|$ ratio increases drastically from 1.9 of CoI$_2$ to 5.9 of Co$_{2/3}$Mg$_{1/3}$I$_2$. To study magnetic structure and phase transitions, we calculated the magnetic structure factor $S(q)$ and the phase diagram. As shown in Fig. \ref{CoMgI}(b), with the suppressed $J_3$ interaction, $S(q)$ is diffusive with soft peak at $M$-point in first and second Brillouin zones. As comparison, the stripe order and zigzag order display sharp features at $M$-point in second and first Brillouin zone respectively, as shown in SM \cite{SM}. Although the residual geometric frustration in Co$_{2/3}$Mg$_{1/3}$I$_2$ somewhat prohibits the QSL state, as shown in Fig. \ref{CoMgI}(c), Co$_{2/3}$Mg$_{1/3}$I$_2$ is already closest to the Kitaev QSL region and would provide a better platform for realizing Kitaev QSL than other candidates such as Na$_2$IrO$_3$ and RuCl$_3$. Then the Kitaev QSL state may readily be realized in Co$_{2/3}$Mg$_{1/3}$I$_2$ with the help of a magnetic field \cite{Gordon2019} or high pressure \cite{Stahl2024}. Indeed, our calculations including a Zeeman term find that the Kitaev QSL phase emerges within the intermediate magnetic field of $2.0\mathrm{-}4.3\; \mathrm{T}$, as seen in Section V of SM \cite{SM}.

In summary, we demonstrate that CoI$_2$ is a potential Kitaev material, using crystal field analysis, DFT calculations, Wannier function analysis, ED and DMRG calculations. Our results show that the high spin Co$^{2+}$ is in the isotropic $J_{\mathrm{eff}}=\frac{1}{2}$ ground state, while the small trigonal distortion induces a weak in-plane anisotropy with $g_{||}=4.62$ (vs $g_{\perp}=3.43$). The long Co-Co distance weakens the $t_{2g}$-$t_{2g}$ hopping and thus reduces the corresponding Heisenberg AFM couplings, but the strong Co-I hybridization enhances the $t_{2g}$-$e_g$ hopping and produces strong FM Kitaev interaction. Both the Kitaev interaction and the geometric frustration from AFM $J_3$ determine the experimental helical magnetic state of CoI$_2$. We propose a partial Mg substitution for Co to suppress the geometric frustration, and indeed find that the resultant Co$_{2/3}$Mg$_{1/3}$I$_2$ has much weakened geometric frustration but the robust bond dependent frustration. As a result, the overwhelming Kitaev interaction renders Co$_{2/3}$Mg$_{1/3}$I$_2$ closest to QSL among several candidate materials. Such prediction may be worth a prompt experimental study.

\section*{Acknowledgements}
This work was supported by National Natural Science Foundation of China (Grants No. 12174062, No. 12241402, and No. 12574127), and by Quantum Science and Technology-National Science and Technology Major Project (2024ZD0300102). Y. Ma and K. Yang contributed equally to this work.


\bibliography{CoI2.bib}

\newpage
\newpage
\clearpage
\begin{appendix}
    \setcounter{figure}{0}
	\setcounter{table}{0}
	\renewcommand{\thefigure}{S\arabic{figure}}
	\renewcommand{\thetable}{S\arabic{table}}
	\renewcommand{\theequation}{S\arabic{equation}}
	\renewcommand{\tablename}{Table}
	\renewcommand{\figurename}{Fig.}

\section*{Supplemental Material for "Kitaev interaction and possible spin liquid state in CoI$_2$ and Co$_{2/3}$Mg$_{1/3}$I$_2$"}

\subsection*{\textbf{I. Crystal field and spin state of CoI$_2$}}

\begin{figure}[H]
	\includegraphics[width=8cm]{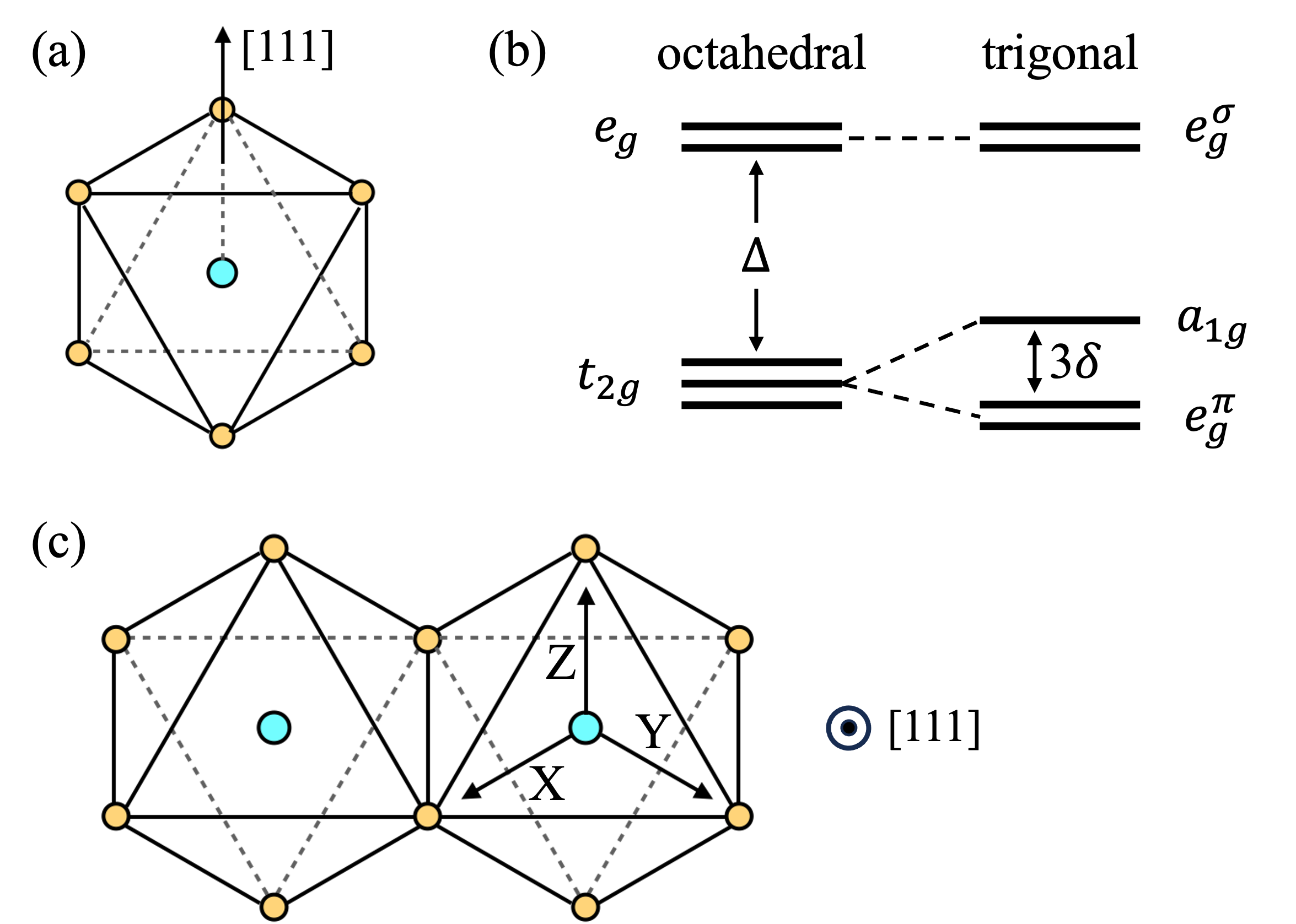}
	\centering
	\caption{(a) Schematic diagram of local CoI$_6$ octahedron. (b) Splitting of energy levels of Co $d$ orbitals under local octahedral crystal field and global trigonal crystal field. Under octahedral crystal field, the degenerate $d$ orbitals split into $t_{2g}$ orbitals and $e_g$ orbitals with energy difference $\Delta$. The trigonal distortion along [111] direction further split $t_{2g}$ orbitals into $e_{g}^\pi$ and $a_{1g}$ orbitals with energy difference $3\delta$. (c) Structure and coordinate of two-site model of Z bond.
	}
  \label{crystal_field}
\end{figure}

\begin{figure*}[t]
	\includegraphics[width=\textwidth]{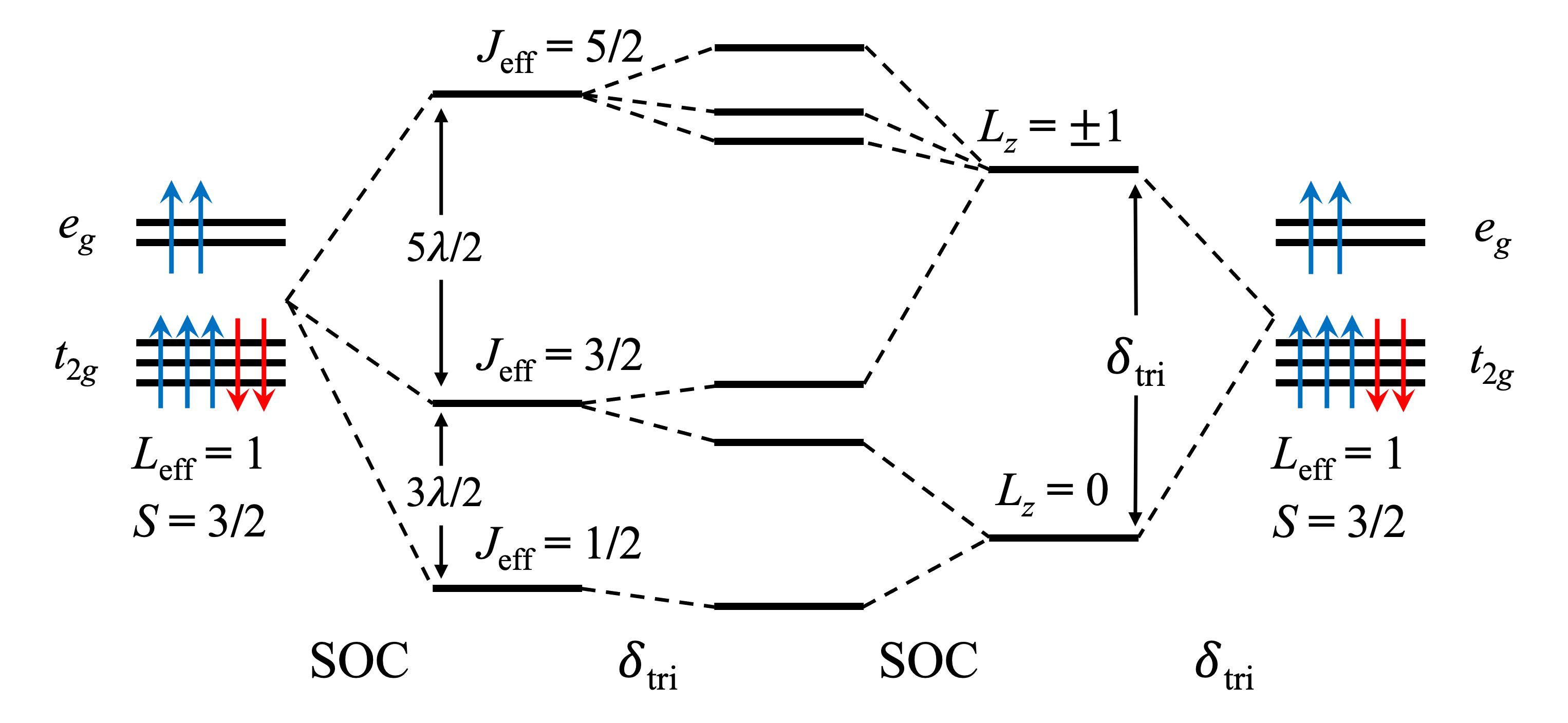}
	\centering
	\caption{Energy splitting of $d^7$ high spin state. The high spin state is 12-fold degenerate states with spin quantum number $S=3/2$ and effective orbital quantum number $L_\mathrm{eff}=1$ on $t_{2g}$ sub-shell. In strong SOC limit, the 12-fold degenerate states split into $J_\mathrm{eff}$ states with quantum number $J_\mathrm{eff}=1/2$, $J_\mathrm{eff}=3/2$ and $J_\mathrm{eff}=5/2$. The trigonal crystal field perturbation would further split $J_\mathrm{eff}$ degenerate states. In strong trigonal crystal field, the $t_{2g}$ states split into $a_{1g}$ and $e_{g}^\pi$ states. For positive trigonal crystal field $\delta>0$, the hole on $t_{2g}$ sub-shell occupies $a_{1g}$ state, leading to $L_z=0$ ground state.
	}
  \label{state}
\end{figure*}

The crystal structure of CoI$_2$ belongs to $P\bar{3}m1$ space group. The edge-sharing CoI$_6$ octahedrons form a triangular network, with the [111] direction of the octahedra aligned along the global z-axis of the van der Waals layer. As shown in Fig. \ref{crystal_field}, the trigonal distortion along the [111] direction of the octahedra splits the degenerate $t_{2g}$ shells into $a_{1g}$ and $e_g^\pi$ sub-shells. For trigonal elongation ($\delta > 0$), the $e_g^\pi$ states lie below the $a_{1g}$ state, whereas for trigonal compression ($\delta < 0$), the $a_{1g}$ state is favored. Through exact diagonalization (ED), we can find the energy difference between $a_{1g}$ and $e_g^\pi$ states is $3\delta$, where $\delta$ is off-diagonal term in crystal field matrix. The spin-orbit coupling (SOC) term, however, has different effect on orbital state. For strong trigonal crystal field limit, the SOC is treated as a perturbation, as illustrated in right part of Fig. \ref{state}. In the $\delta>0$ crystal field, the two minor spins occupy the lower $e_g^\pi$ orbitals, resulting in a quenched orbital moment ground state ($L_z = 0$). However, in strong SOC limit, as illustrated in left part of Fig. \ref{state}, the high spin state splits into $J_\mathrm{eff}$ states, and the trigonal crystal field is treated as perturbation to further split $J_\mathrm{eff}$ states. The wavefunctions of 12 $J_\mathrm{eff}$ states are listed in Eq. S1-S12. The left side of equals sign represents $J_\mathrm{eff}$ state $\left|J, j_z\right\rangle$, while the right side represents $\left|S_z, L_z\right\rangle$. To study $d^7$ $J_\mathrm{eff}$ states, we must include multi-electron effect in our model with help of ED, as for example the $\left|S_z, L_z\right\rangle=\left|\frac12,0\right\rangle$ and the $\left|S_z, L_z\right\rangle=\left|\frac32,0\right\rangle$ are in-distinguishable in density functional theory (DFT) calculation.

\begin{equation}
	\left|\frac12,-\frac12\right\rangle =
	  \frac{1}{\sqrt6}\left|\frac12,-1\right\rangle
	 -\frac{1}{\sqrt3}\left|-\frac12,0\right\rangle
	 +\frac{1}{\sqrt2}\left|-\frac32,1\right\rangle
\end{equation}

\begin{equation}
	\left|\frac12,\frac12\right\rangle =
	  \frac{1}{\sqrt2}\left|\frac32,-1\right\rangle
	 -\frac{1}{\sqrt3}\left|\frac12,0\right\rangle
	 +\frac{1}{\sqrt6}\left|-\frac12,1\right\rangle
\end{equation}

\begin{equation}
	\left|\frac32,-\frac32\right\rangle =
	 -\sqrt{\frac25}\left|-\frac12,-1\right\rangle
	 +\sqrt{\frac35}\left|-\frac32,0\right\rangle
\end{equation}

\begin{equation}
	\left|\frac32,-\frac12\right\rangle =
	 -\sqrt{\frac{8}{15}}\left|\frac12,-1\right\rangle
	 +\sqrt{\frac{1}{15}}\left|-\frac12,0\right\rangle
	 +\sqrt{\frac{2}{5}}\left|-\frac32,1\right\rangle
\end{equation}

\begin{equation}
	\left|\frac32,\frac12\right\rangle =
	 -\sqrt{\frac25}\left|\frac32,-1\right\rangle
	 -\sqrt{\frac{1}{15}}\left|\frac12,0\right\rangle
	 +\sqrt{\frac{8}{15}}\left|-\frac12,1\right\rangle
\end{equation}

\begin{equation}
	\left|\frac32,\frac32\right\rangle =
	 -\sqrt{\frac35}\left|\frac32,0\right\rangle
	 +\sqrt{\frac25}\left|\frac12,1\right\rangle
\end{equation}

\begin{equation}
	\left|\frac52,-\frac52\right\rangle =
	 \left|-\frac32,-1\right\rangle
\end{equation}

\begin{equation}
	\left|\frac52,-\frac32\right\rangle =
	  \sqrt{\frac35}\left|-\frac12,-1\right\rangle
	 +\sqrt{\frac25}\left|-\frac32,0\right\rangle
\end{equation}

\begin{equation}
	\left|\frac52,-\frac12\right\rangle =
	  \sqrt{\frac{3}{10}}\left|\frac12,-1\right\rangle
	 +\sqrt{\frac35}\left|-\frac12,0\right\rangle
	 +\sqrt{\frac{1}{10}}\left|-\frac32,1\right\rangle
\end{equation}

\begin{equation}
	\left|\frac52,\frac12\right\rangle =
	 \sqrt{\frac{1}{10}}\left|\frac32,-1\right\rangle
	 +\sqrt{\frac35}\left|\frac12,0\right\rangle
	 +\sqrt{\frac{3}{10}}\left|-\frac12,1\right\rangle
\end{equation}

\begin{equation}
	\left|\frac52,\frac32\right\rangle =
	  \sqrt{\frac25}\left|\frac32,0\right\rangle
	 +\sqrt{\frac35}\left|\frac12,1\right\rangle
\end{equation}

\begin{equation}
	\left|\frac52,\frac52\right\rangle =
	 \left|\frac32,1\right\rangle
\end{equation}

\subsection*{\textbf{II. Occupation and magnetic moment}}

\begin{figure}
	\includegraphics[width=8cm]{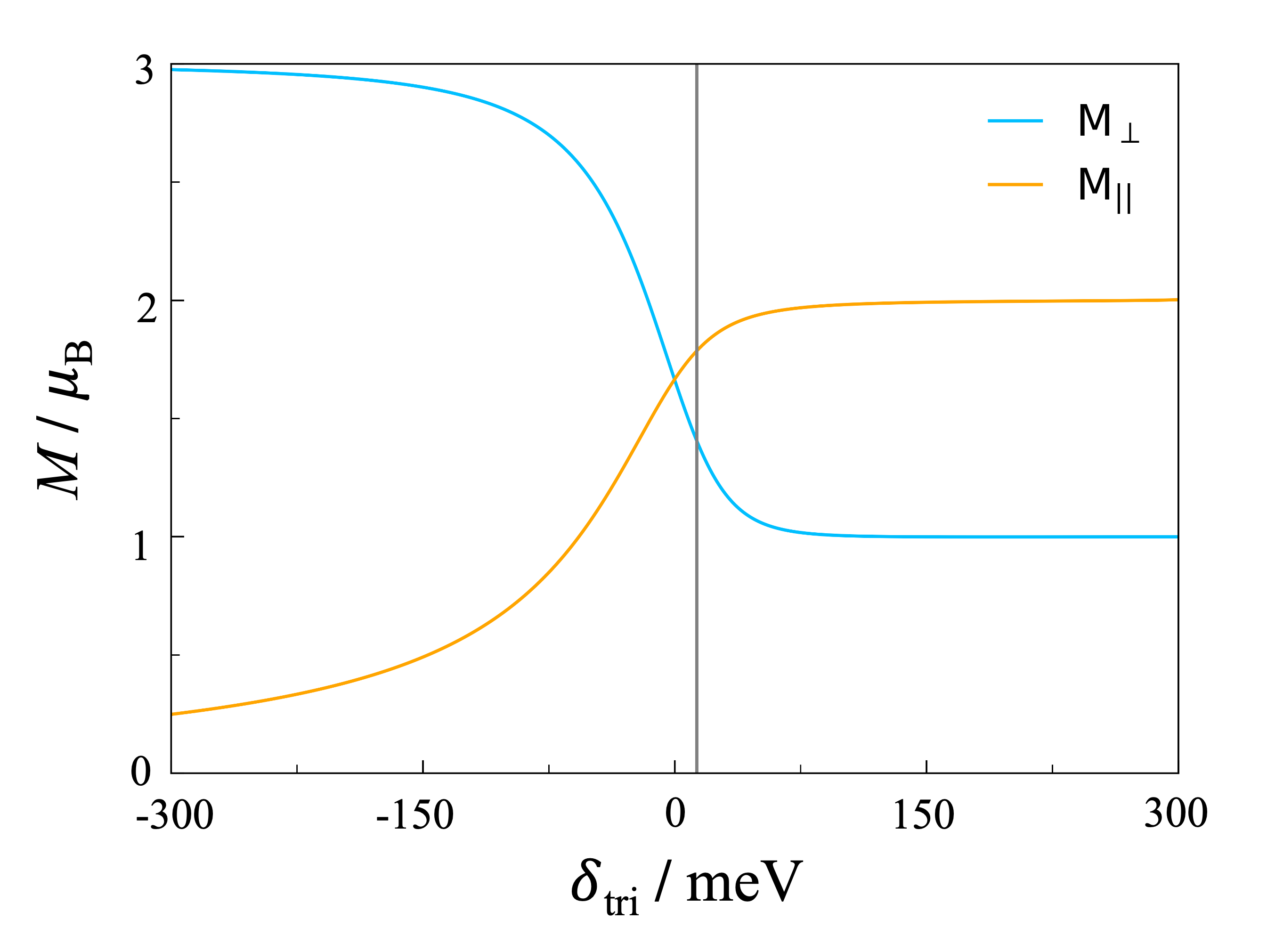}
	\centering
	\caption{Spin magnetic moment under different trigonal crystal field. The out-of-plane magnetic moment is denoted by blue curve while the in-plane magnetic moment is denoted by yellow curve. The gray line denotes realistic trigonal crystal field strength $\delta_{\rm {tri}} = 3\delta=13.2\;\mathrm{meV}$ of CoI$_2$.
	}
  \label{spin_m}
  \end{figure}

  \begin{figure}
	\includegraphics[width=8cm]{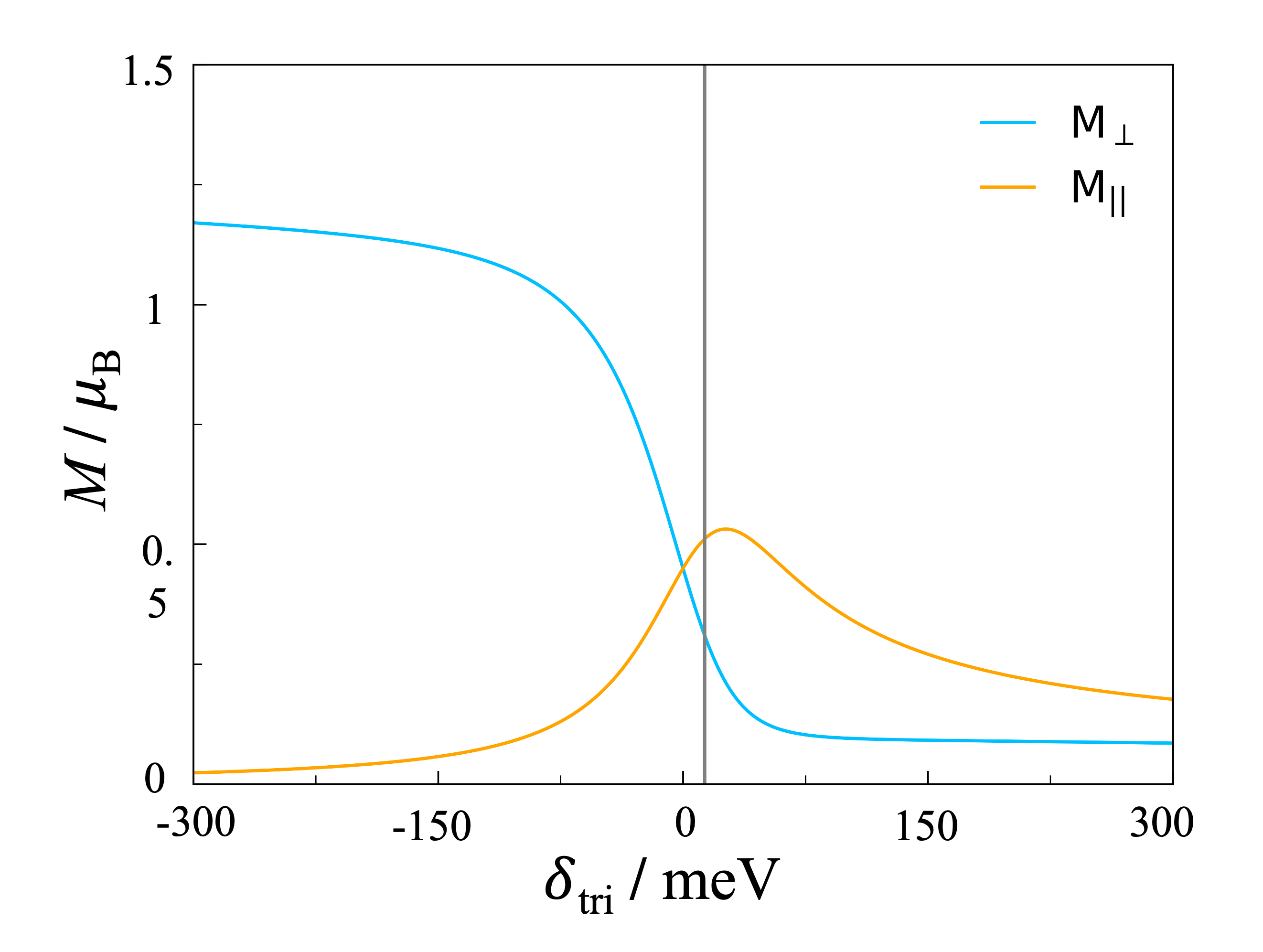}
	\centering
	\caption{Orbital magnetic moment under different trigonal crystal field. The out-of-plane magnetic moment is denoted by blue curve while the in-plane magnetic moment is denoted by yellow curve. The gray line denotes realistic trigonal crystal field strength $\delta_{\rm {tri}} = 3\delta=13.2\;\mathrm{meV}$ of CoI$_2$.
	}
  \label{orbital_m}
  \end{figure}

The occupation and magnetic moment are calculated with ED method on one-site model. The one-site model contains crystal field term, Coulomb repulsion and SOC,

\begin{equation}
\label{eq:one-site}
	H_{\mathrm{one}} = H_{\mathrm{CF}} + H_{U} + H_{\mathrm{SOC}}.
\end{equation}

We approximate the Coulomb repulsion using the isotropic Kanamori form, with the intra-orbital Hubbard parameter $U$ and Hund's coupling $J_{\mathrm{H}}$. The inter-orbital Hubbard parameter $U'$ is taken as $U' = U - 2J_{\mathrm{H}}$ in its spherically symmetric form \cite{Georges2013}. The Coulomb repulsion term is written as
  \begin{equation}
  \begin{aligned}
  H_U= & U \sum_{i, a} n_{i, a, \uparrow} n_{i, a, \downarrow}+\left(U^{\prime}-J_{\mathrm{H}}\right) \sum_{i, a<b, \sigma} n_{i, a, \sigma} n_{i, b, \sigma} \\
  & +U^{\prime} \sum_{i, a \neq b} n_{i, a, \uparrow} n_{i, b, \downarrow}-J_{\mathrm{H}} \sum_{i, a \neq b} c_{i, a \uparrow}^{\dagger} c_{i, a \downarrow} c_{i, b \downarrow}^{\dagger} c_{i, b \uparrow} \\
  & +J_{\mathrm{H}} \sum_{i, a \neq b} c_{i, a \uparrow}^{\dagger} c_{i, a \downarrow}^{\dagger} c_{i, b \downarrow} c_{i, b \uparrow},
  \end{aligned}
  \end{equation}
  where a and b are orbitals and $\sigma$ and $\sigma^\prime$ are spin states. We take intra-orbital Hubbard $U=4.0 \; \mathrm{eV}$ and Hund's coupling $J_{\mathrm{H}} = 0.9 \; \mathrm{eV}$ as used in DFT calculation.

  The SOC term $H_{\mathrm{SOC}} = \lambda L\cdot S$, is written as single-particle form in calculation:
  \begin{equation}
	  H_{\mathrm{SOC}}=\zeta\sum_{j}\mathbf{l}_j\cdot\sigma_j
  \end{equation}
  where $\zeta=3\lambda=0.08$ eV is the atomic SOC strength and $j$ represents the holes. The one-site model is solved with ED method, and the ground state is projected onto $J_\mathrm{eff}=1/2$, $L_z=0$ and $L_z=\pm1$ states to obtain the occupation $\left|\langle\Psi_\mathrm{ground}|J_\mathrm{eff}=1/2\rangle\right|^2$, $\left|\langle\Psi_\mathrm{ground}|L_z=0\rangle\right|^2$ and $\left|\langle\Psi_\mathrm{ground}|L_z=\pm1\rangle\right|^2$. The magnetic moment is obtained by solving one-site model under a weak out-of-plane or in-plane Zeeman field. The ground state is used to calculate the expectation of total magnetic moment $\langle(\vec{L}+2\vec{S})\rangle$, spin magnetic moment $\langle 2\vec{S}\rangle$ and orbital magnetic moment $\langle \vec{L}\rangle$.

The spin magnetic moment, orbital magnetic moment and total magnetic moment are shown in Fig. \ref{spin_m}, Fig. \ref{orbital_m} and inset figure in Fig. 1(c). In the trigonal compression limit (left part of the figure), the ground state is characterized by $L_z = \pm 1$. Because of unquenched orbital magnetic moment in $L_z=\pm1$ orbitals, the spin is constrained along z-axis, resulting in a negligible in-plane spin magnetic moment. Note that the orbital magnetic moment exceeding 1 $\mu_\mathrm{B}$ is caused by the finite crystal field $\Delta$, leading to a non-zero occupation of the $e_g^\sigma$ orbitals. In the trigonal elongation limit (right part of the figure), the ground state corresponds to $L_z = 0$. The orbital magnetic moment is quenched, leading to weak in-plane anisotropy. Consequently, the spin magnetic moment is primarily oriented in the in-plane direction. For SOC limit (middle part of the figure), the ground state is $J_\mathrm{eff}=1/2$ state. The $J_\mathrm{eff}=1/2$ is isotropic state, resulting in the same magnetic moment along any directions.

\subsection*{\textbf{III. Wannier results}}

\begin{figure}
	\includegraphics[width=8cm]{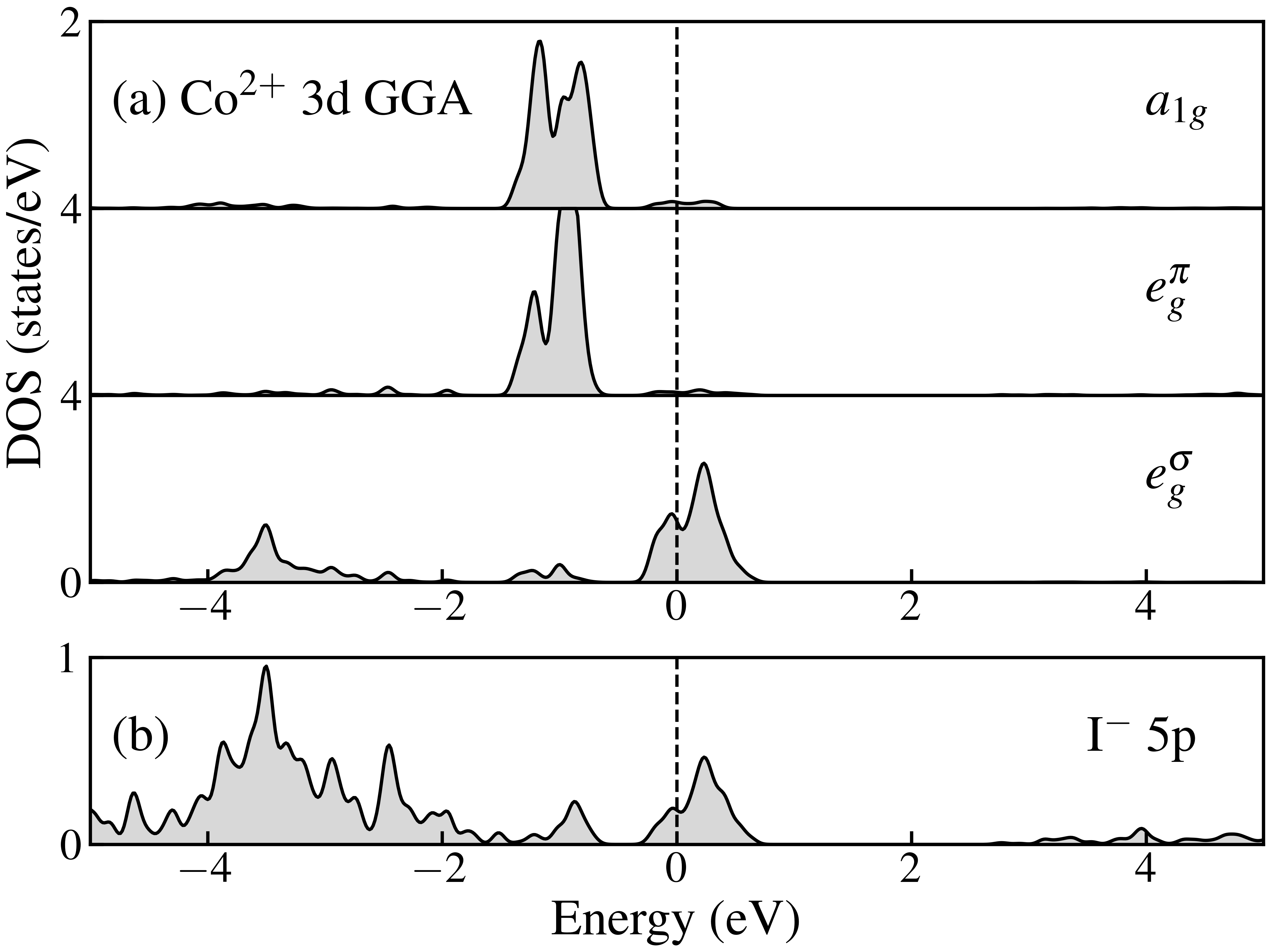}
	\centering
	\caption{(a) Co 3$d$ and (b) I 5$p$ density of states (DOS) from GGA calculation. The Fermi level is set at zero energy. The DOS of Co are projected to $a_{1g}$, $e_g^\pi$ and $e_g^{\sigma}$ orbitals, considering the triangular crystal field.
	}
  \label{dos_gga}
  \end{figure}

The Wannier function analysis is based on density functional theory (DFT) calculation. The DFT calculations are performed using Vienna Ab Initio Simulation Package (VASP) \cite{VASP}, within the framework of generalized gradient approximation (GGA) \cite{GGA}. The kinetic energy cutoff of 500 eV and a Gamma centered $k$-mesh of 9$\times$9$\times$4 are used for the unit cell. We use the $P\bar{3}m1$ crystal structure of CoI$_2$ \cite{McGuire2017}. And the lattice constants are optimized to be $a=b=$ 3.87 $\mathrm{\AA}$ and $c=$ 7.06 $\mathrm{\AA}$. In the structure optimization, all atoms are fully relaxed with the tolerance of $10^{-6}$ eV for total energy and $0.01$ eV/$\mathrm{\AA}$ for atomic forces. As seen in Fig. \ref{dos_gga}, the $a_{1g}$ and $e_g^\pi$ states are nearly degenerate, suggesting a small trigonal crystal field.

\begin{figure}
	\includegraphics[width=8cm]{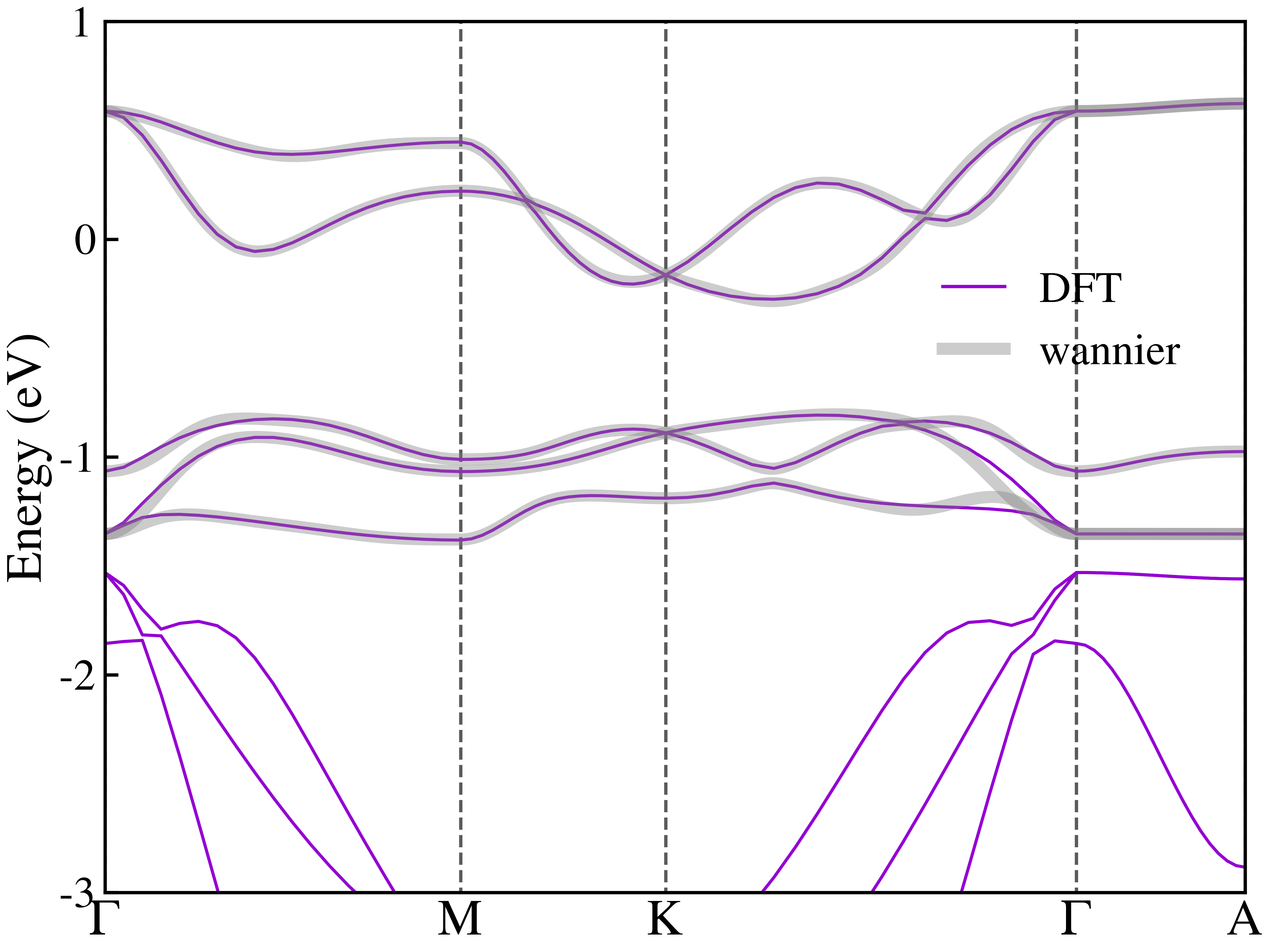}
	\centering
	\caption{Band structure from GGA calculation. The purple line and gray line represent DFT band structure and Wannier band structure respectively.
	}
  \label{wannier}
  \end{figure}

\begin{figure}
	\includegraphics[width=8cm]{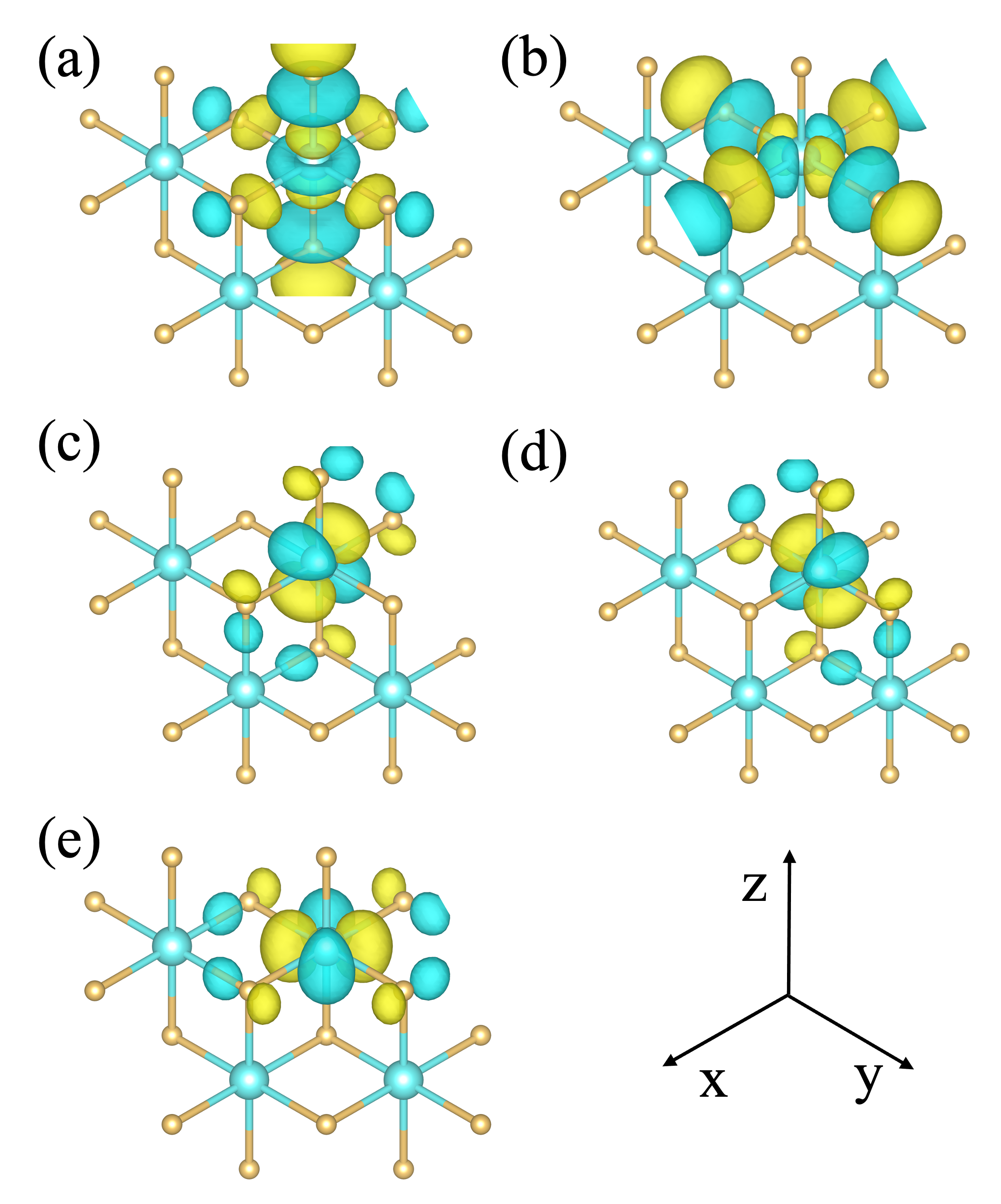}
	\centering
	\caption{Contour-surface plots of Co (a) $d_{3z^2-r^2}$, (b) $d_{x^2-y^2}$, (c) $d_{xz}$, (d) $d_{yz}$ and (e) $d_{xy}$ Wannier functions.
	}
  \label{wannier_functions}
  \end{figure}

We perform Wannier function analysis based on GGA result, using Wannier90 package \cite{wannier90}. To study magnetic properties of Co, we construct Wannier function based on Co $d$ orbitals. As shown in Fig. \ref{wannier}, the Wannier functions band structure well reproduce the DFT band structure. To further check the reliability of Wannier function analysis, we plot Wannier function in Fig. \ref{wannier_functions}. These Wannier functions are localized at Co ions with contribution from $p$ orbitals of ligands, which results from $p$-$d$ bonds and through which the indirect hopping processes are included in our model. Note that the hybridization between the $e_g^\sigma$ orbitals and $p$ orbitals is particularly strong, leading to significant $e_g$-$t_{2g}$ hopping terms.

  The crystal field term $H_\mathrm{CF}$ and hopping matrix $H_t^{(ij)}$ are obtained from Wannier function analysis. Here, we present the hopping matrices for the $Z$ bond. The hopping matrices of $X$ and $Y$ bonds are related with $Z$ bond matrices by $C_3$ rotation in the [111] direction. For CoI$_2$, the crystal field term $H_\mathrm{CF}$ and hopping matrices between first nearest neighbors $H_t^{1st}$, second nearest neighbors $H_t^{2nd}$ and third nearest neighbors $H_t^{3rd}$ are (in meV)

\begin{equation}
    H_\mathrm{CF}=
    \begin{pmatrix}
      5280.0&0		&-2.1	&-2.1	&4.1\\
      0		&5280.0&-3.6	&3.6 	&0\\
      -2.1	&-3.6	&4258.6&4.4	&4.4\\
      -2.1	&3.6 	&4.4 	&4258.6&4.4\\
      4.1	    &0		&4.4 	&4.4 	&4258.6
    \end{pmatrix},
\end{equation}

\begin{equation}
    H_t^{1st}=
    \begin{pmatrix}
      18.1	&0		&10.3	&10.3	&177.1\\
      0		&22.6	&5.3 	&-5.3	&0\\
      10.3	&5.3 	&29.2	&24.7	&5.1\\
      10.3	&-5.3	&24.7	&29.2	&5.1\\
      177.1 &0		&5.1 	&5.1 	&-44.1
    \end{pmatrix},
\end{equation}

\begin{equation}
    H_t^{2nd}=
    \begin{pmatrix}
      -14.0	&0		&-11.8	&-11.8	&39.6\\
      0		&7.8		&2.9		&-2.9	&0\\
      -11.8	&2.9		&-0.7	&-39.1	&3.2\\
      -11.8	&-2.9	&-39.1	&-0.7	&3.2\\
      39.6	&0		&3.2		&3.2		&2.6
    \end{pmatrix},
\end{equation}

\begin{equation}
    H_t^{3rd}=
    \begin{pmatrix}
      -44.0	&0		&-5.8	&-5.8	&20.9\\
      0		&92.0	&-4.5	&4.5		&0\\
      -5.8	&-4.5	&8.1		&-5.3	&5.6\\
      -5.8	&4.5		&-5.3	&8.1		&5.6\\
      20.9	&0		&5.6		&5.6		&-32.7
    \end{pmatrix}.
\end{equation}

For Co$_{2/3}$Mg$_{1/3}$I$_2$, the crystal field term $H_\mathrm{CF}$ and hopping matrices between first nearest neighbors $H_t^{1st}$, second nearest neighbors $H_t^{2nd}$ and third nearest neighbors $H_t^{3rd}$ are (in meV)

\begin{equation}
    H_\mathrm{CF}=
    \begin{pmatrix}
      4659.8&0		&4.6 	&4.6 	&-9.2\\
      0		&4659.8&8.0	&-8.0	&0\\
      4.6 	&8.0		&3671.2&1.5	&1.5\\
      4.6 	&-8.0	&1.5 	&3671.2&1.5\\
      -9.2	&0		&1.5		&1.5		&3671.2
    \end{pmatrix},
\end{equation}

\begin{equation}
    H_t^{1st}=
    \begin{pmatrix}
      12.2	&0		&9.9 	&9.9		&148.0\\
      0		&58.8	&0.3		&-0.3	&0\\
      9.9		&0.3		&17.4	&32.0	&-5.1\\
      9.9		&-0.3	&32.0	&17.4	&-5.1\\
      148.0	&0		&-5.1	&-5.1	&-2.7
    \end{pmatrix},
\end{equation}

\begin{equation}
    H_t^{2rd}=
    \begin{pmatrix}
      -1.6	&-4.7	&-11.7	&-13.4	&38.0\\
      4.7		&2.8 	&-1.5	&-0.1	&7.3\\
      -13.4	&0.1 	&-1.7	&-41.4	&7.7\\
      -11.7	&1.5 	&-29.2	&-1.7	&0.1\\
      38.0	&-7.3	&0.1		&7.7		&-0.3
    \end{pmatrix},
\end{equation}

\begin{equation}
    H_t^{3rd}=
    \begin{pmatrix}
      -25.2	&0		&-6.6	&-6.6	&19.3\\
      0		&47.9	&-3.8	&3.8 	&0\\
      -6.6	&-3.8	&4.9 	&-7.1	&9.4\\
      -6.6	&3.8		&-7.1	&4.9 	&9.4\\
      19.3	&0		&9.4		&9.4		&-31.4
    \end{pmatrix}.
\end{equation}

\subsection*{\textbf{IV. Exchange parameters}}

We obtained the exchange parameters by solving tow-site Hubbard model,
\begin{equation}
\label{eq:Hubbard}
	H = H_{\mathrm{one}} + H_{t},
\end{equation}
where the crystal field terms and hopping terms are obtained through Wannier function analysis.

We used exact diagonalization and projection to obtain the exchange parameters. Technically, our approach begins with the exact diagonalization of the Hubbard model to obtain the four lowest-energy eigenstates, denoted as $|n\rangle$, along with their corresponding energy spectrum. We then define a target subspace spanned by the pure $J_\mathrm{eff}=1/2$ basis states, $|n_l\rangle$. To derive the effective Hamiltonian, the low-energy eigenstates $|n\rangle$ are first projected onto the subspace spanned by $|n_l\rangle$. Subsequently, we apply Löwdin symmetric orthonormalization to construct a set of orthonormal basis states, and at the same time the diagonalized energy spectrum is transformed into effective Hamiltonian. This exact diagonalization and projection scheme is identical to the methodology employed for the calculations of Na$_2$IrO$_3$ and RuCl$_3$ \cite{Winter2016}.

The Hubbard model is solved by ED method and the obtained four lowest eigenstates are $|J_\mathrm{eff}=\frac{1}{2}, J_\mathrm{eff}=\frac{1}{2}\rangle$, $|J_\mathrm{eff}=\frac{1}{2}, J_\mathrm{eff}=-\frac{1}{2}\rangle$, $|J_\mathrm{eff}=-\frac{1}{2}, J_\mathrm{eff}=\frac{1}{2}\rangle$ and $|J_\mathrm{eff}=-\frac{1}{2}, J_\mathrm{eff}=-\frac{1}{2}\rangle$. Due to the limited trigonal crystal field, the $J_\mathrm{eff}=\frac{1}{2}$ eigenstates are mixed with $J_\mathrm{eff}=\frac{3}{2}$ and $J_\mathrm{eff}=\frac{5}{2}$ states, as shown in Fig. \ref{state}. As we show in main texture, the occupation of $J_\mathrm{eff}=\frac{1}{2}$ is over 0.98 and the majority of eigenstate is still $J_\mathrm{eff}=\frac{1}{2}$ state.

\begin{figure} [H]
	\includegraphics[width=8cm]{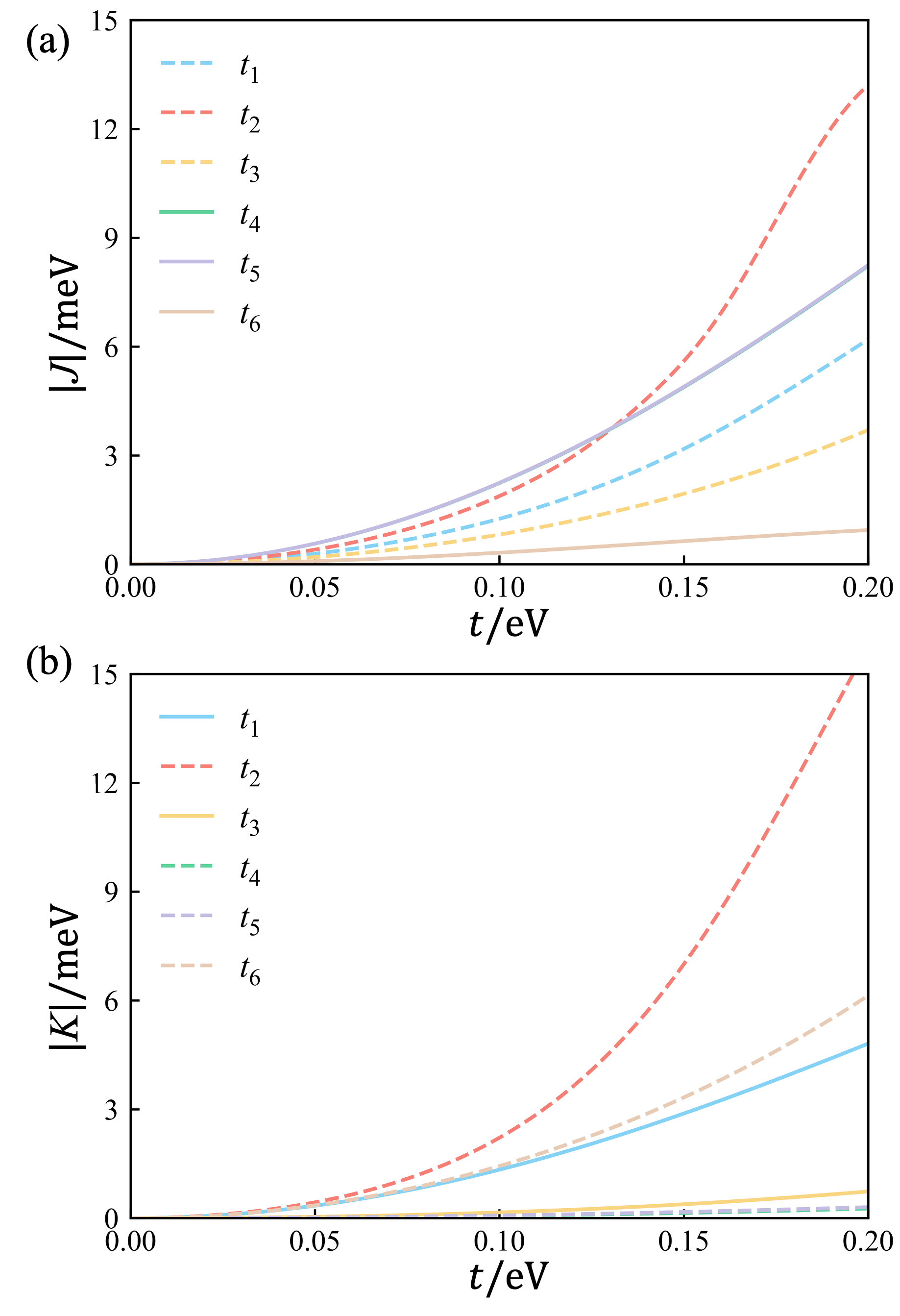}
	\centering
	\caption{ The dependence of the exchange interaction on the hopping terms $t_{1-6}$: (a) Heisenberg interaction $J$, (b) Kitaev interaction $K$. The dash line and solid line indicate FM interaction and AFM interaction, respectively.
	}
  \label{JKt}
  \end{figure}

To study contributions from each channel, we calculated the dependence of Heisenberg interaction $J$ and Kitaev interaction $K$ on each hopping channel, as shown in Fig. \ref{JKt}. For first nearest neighbors, the strongest channel is $t_{2g}$-$e_g$ hopping term $t_6=177.1\;\mathrm{meV}$. And the single channel contributions in CoI$_2$ are listed in Table \ref{contribution}. The $t_6$ channel contributes a AFM Heisenberg interaction $J=0.81\;\mathrm{meV}$ and a FM interaction $K=-4.73\;\mathrm{meV}$. In the meantime, the $t_3=-44.1\;\mathrm{meV}$ channel contributes a AFM Kitaev interaction $K=0.03\;\mathrm{meV}$ and a FM Heisenberg interaction $J=-0.15\;\mathrm{meV}$. For second nearest neighbors, the $t_{2g}$-$t_{2g}$ hopping term $t_2=-39.1\; \mathrm{meV}$ contributes a FM Heisenberg interaction $J=-0.24\; \mathrm{meV}$ and a FM Kitaev interaction $K=-0.26\; \mathrm{meV}$. And the $t_{2g}$-$e_g$ hopping term $t_6=39.6\;\mathrm{meV}$ contributes a AFM Heisenberg interaction $J=0.05\;\mathrm{meV}$ and a FM Kitaev interaction $K=-0.22\;\mathrm{meV}$. For third nearest neighbors, the strongest hopping channel is $e_g$-$e_g$ hopping channel $t_4=92.0\;\mathrm{meV}$. The $t_4$ channel contributes a AFM Heisenberg interaction $J=1.9\;\mathrm{meV}$ and a FM Kitaev interaction $K=-0.05\;\mathrm{meV}$. The $t_4$ channel results the AFM Heisenberg interaction between third nearest neighbors.

\begin{table}[t]
\centering
\setlength{\tabcolsep}{3.2mm}
\caption{Magnetic interactions (meV) from each single channel for CoI$_2$ between first, second and third nearest neighbors.}
\label{contribution}
\begin{tabular}{cccccc}
  \hline \hline
&   &  $J$            &  $K$             & $\Gamma$       &  $\Gamma^\prime$ \\ \hline
\multirow{6}{*}{1NN} &$t_1$ & -0.10 & 0.12  & -0.01  & 0.01 \\
                     &$t_2$ & -0.10 & -0.10 & -0.00  & -0.03 \\
                     &$t_3$ & -0.15 & 0.03  & -0.03  & -0.02 \\
                     &$t_4$ &  0.12 & -0.00  & 0.02  &  0.01 \\
                     &$t_5$ &  0.08 & -0.00  & 0.01  &  0.01 \\
                     &$t_6$ &  0.81 & -4.73  & 0.17  & -0.22 \\ \hline
\multirow{6}{*}{2NN} &$t_1$ & -0.00 & 0.00  & -0.00  & 0.00 \\
                     &$t_2$ & -0.24 & -0.26 & -0.01  & -0.08 \\
                     &$t_3$ & -0.00 & 0.00  & -0.00  & -0.00 \\
                     &$t_4$ &  0.01 & -0.00  & 0.00  &  0.00 \\
                     &$t_5$ &  0.05 & -0.00  & 0.01  &  0.01 \\
                     &$t_6$ &  0.05 & -0.22  & 0.01  & -0.01 \\ \hline
\multirow{6}{*}{3NN} &$t_1$ & -0.01 & 0.01  & -0.00  & 0.00 \\
                     &$t_2$ & -0.00 & -0.00 & -0.00  & -0.00 \\
                     &$t_3$ & -0.08 & 0.02  & -0.01  & -0.00 \\
                     &$t_4$ &  1.90 & -0.05  & 0.34  &  0.26 \\
                     &$t_5$ &  0.44 & -0.01  & 0.07  &  0.06 \\
                     &$t_6$ &  0.02 & -0.06  & 0.00  & -0.00 \\ \hline
\hline
\end{tabular}
\end{table}

\subsection*{\textbf{V. Magnetic structure}}

We calculate the ground state of the spin model using density matrix renormalization group (DMRG) method \cite{dmrg} implemented in the ITensor library \cite{Fishman2022} for triangular lattice and ED method for hexagonal lattice. For triangular lattice, we performed DMRG calculations on a $6 \times 6$ lattice with periodic boundary conditions, performing over 10 sweeps with up to 1600 states in each DMRG block. For hexagonal lattice, we used ED method to calculate the ground state on a $6 \times 4$ lattice.

\begin{figure}
	\includegraphics[width=8cm]{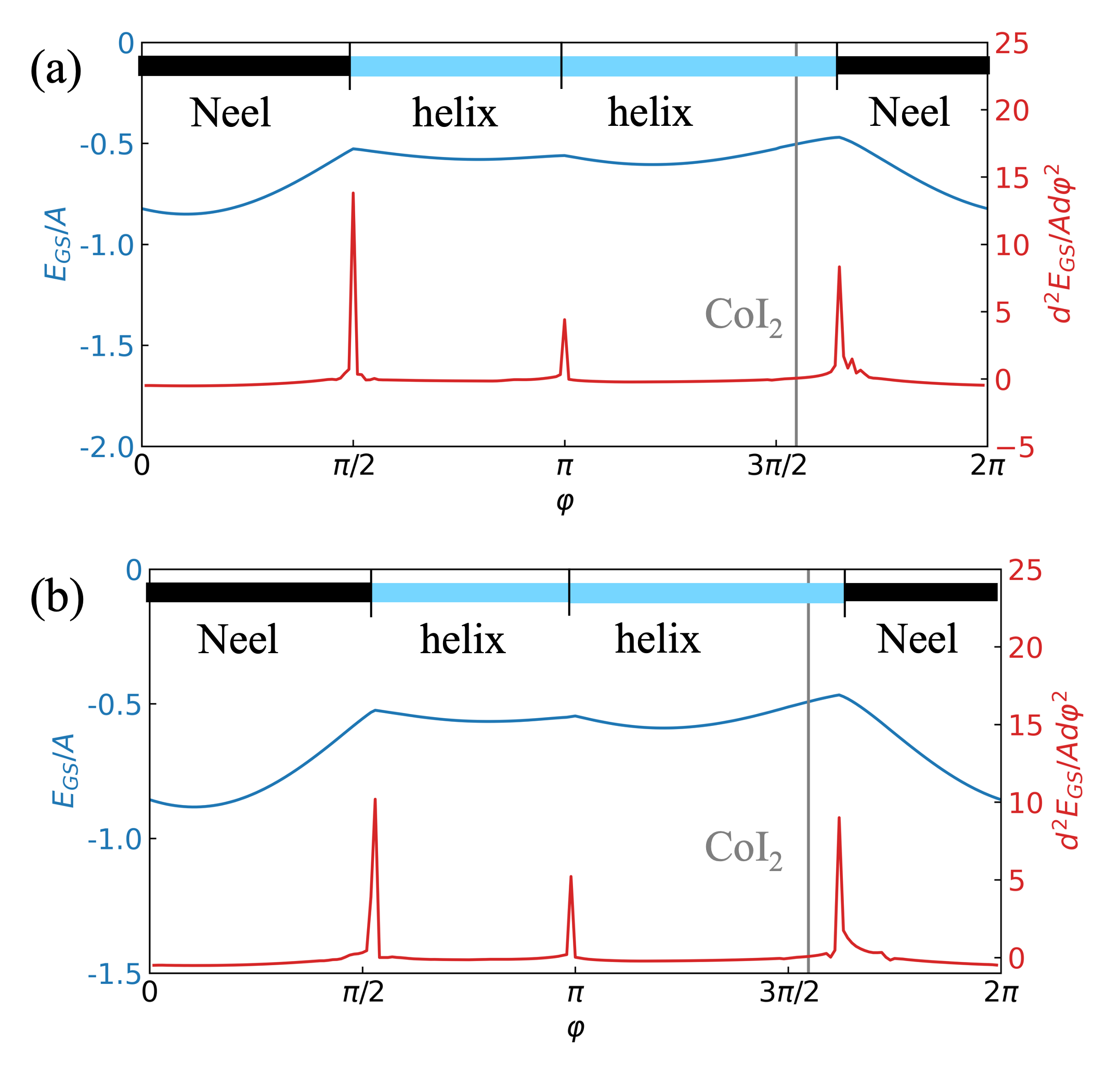}
	\centering
	\caption{ Phase diagram of (a) $J_1$-$K_1$-$J_3$ model and (b) $J_1$-$K_1$-$J_2$-$J_3$ model.
	}
  \label{J2_phase}
  \end{figure}

Here we also calculated the phase diagram for the triangular lattice with the model containing interactions between first, second and third nearest neighbors. As shown in Fig. \ref{J2_phase}, the phase diagram is almost identical to the one without the second nearest neighbor interaction, due to the very weak $J_2$ interaction. Therefore, in our calculation of geometric frustration, only $J_3$ is considered significant.

\begin{figure}
	\includegraphics[width=8cm]{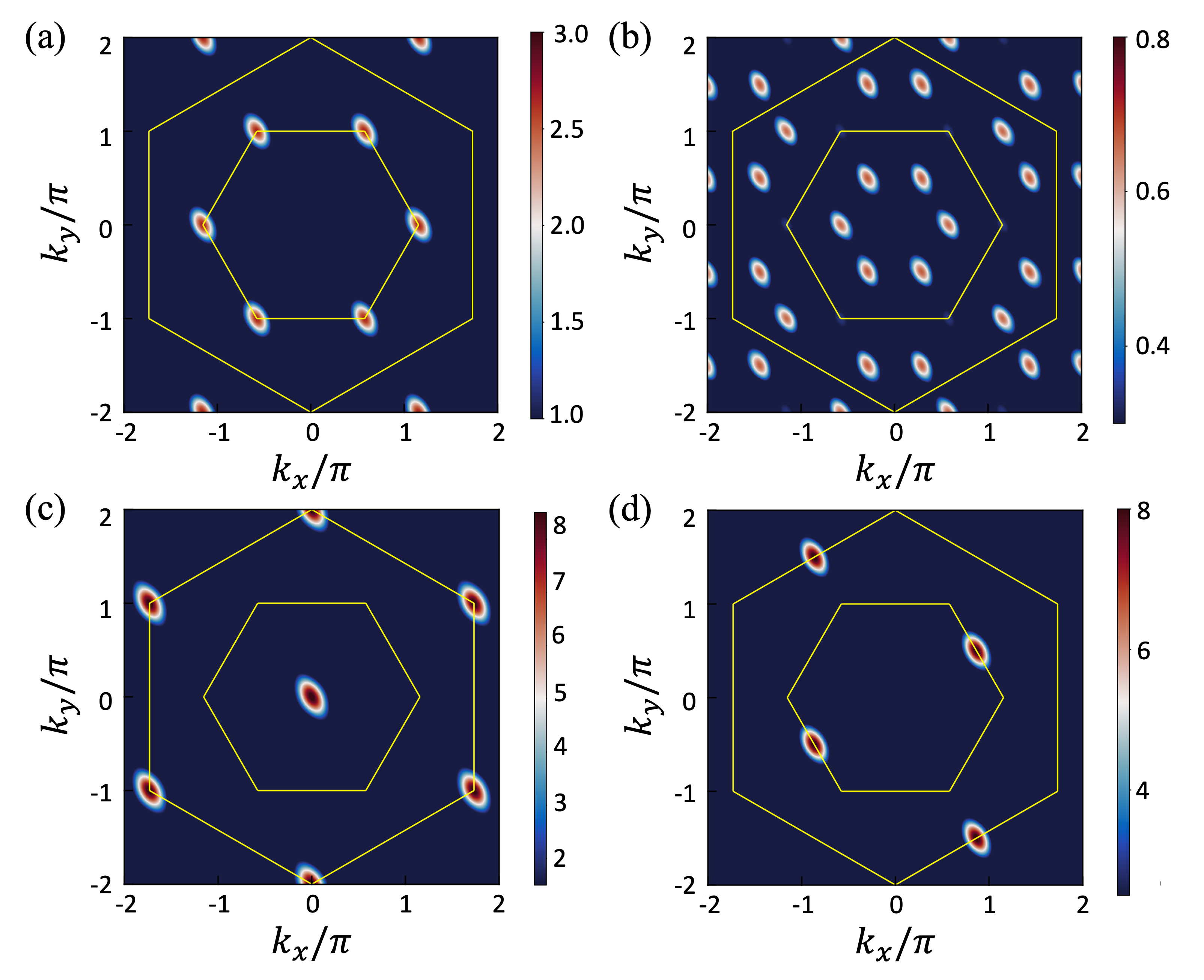}
	\centering
	\caption{Magnetic structure factor $S(q)$ of (a) 120$^\circ$ Neel AFM state, (b) helical AFM state, (c) FM state and (d) stripe AFM state.
	}
  \label{triangular_magnetic_structure}
  \end{figure}

To determine the ground state, we calculated the magnetic structure factor $S(q)$,
\begin{equation}
    S(q)=\frac{1}{N}\sum_{ij}S_{i}S_{j}e^{i\vec{q}\cdot\vec{R}_{j-i}},
\end{equation}
where $S_i$ and $S_j$ are spin moment expectation of site $i$ and site $j$, $N$ is the number of sites and $\vec{R}_{j-i}$ is displacement between site $i$ and site $j$ in real space. The magnetic structure factor of triangular lattice is shown in Fig. \ref{triangular_magnetic_structure}.

\begin{figure}
	\includegraphics[width=8cm]{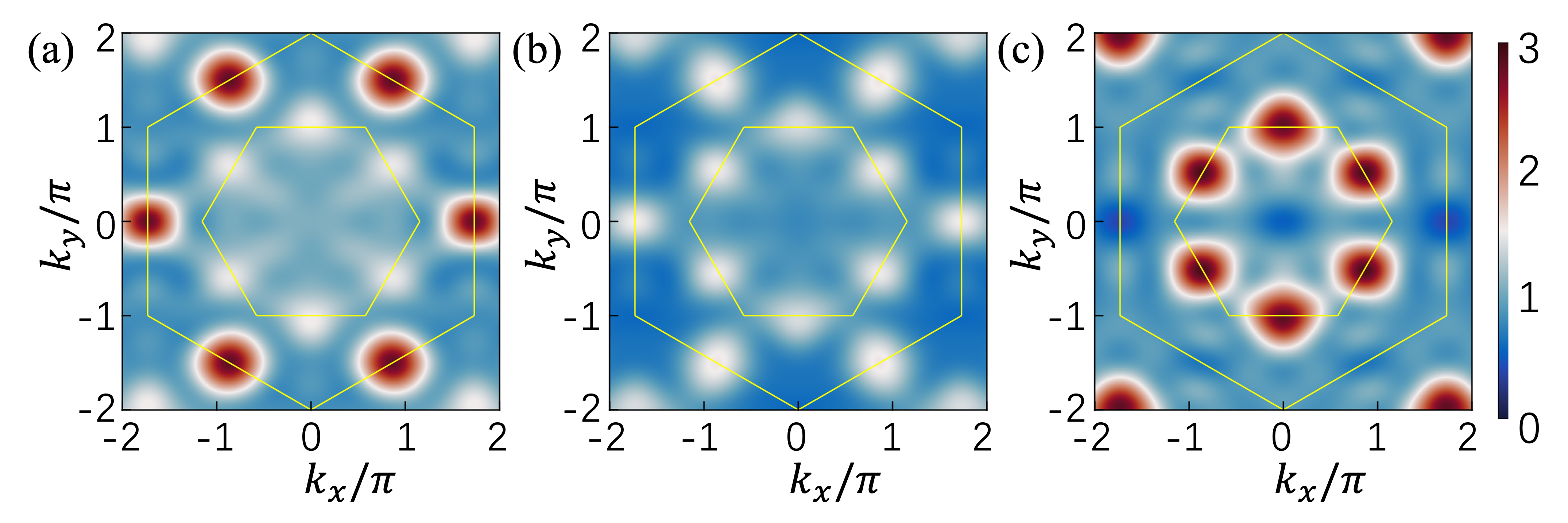}
	\centering
	\caption{Magnetic structure factor $S(q)$ of Co$_{2/3}$Mg$_{1/3}$I$_2$ with (a) $0\%$ $J_3$, (b) $30\%$ $J_3$, (c) $100\%$ $J_3$. The colormap is limited to the range of 0 to the maximum value $S(q)_\mathrm{max}$ of zigzag and stripe order.
	}
  \label{sq_CoMgI}
  \end{figure}

The magnetic structure factor $S(q)$ of Co$_{2/3}$Mg$_{1/3}$I$_2$ displays the sequence of phase transition associated with $J_3$ interaction. As shown in Fig. \ref{sq_CoMgI}(a), without $J_3$, $S(q)$ displays sharp peak at $M$-point in second Brillouin zone, which corresponds to stripe AFM order. In Fig. \ref{sq_CoMgI}(b), a limited $J_3$ induces diffused $S(q)$ with soft peak at $M$-point in first and second Brillouin zone. And as shown in Fig. \ref{sq_CoMgI}(c), where $J_3$ is relatively strong, $S(q)$ displays sharp peak at $M$-point in first Brillouin zone and the ground state is zigzag AFM.

\begin{figure}
	\includegraphics[width=8cm]{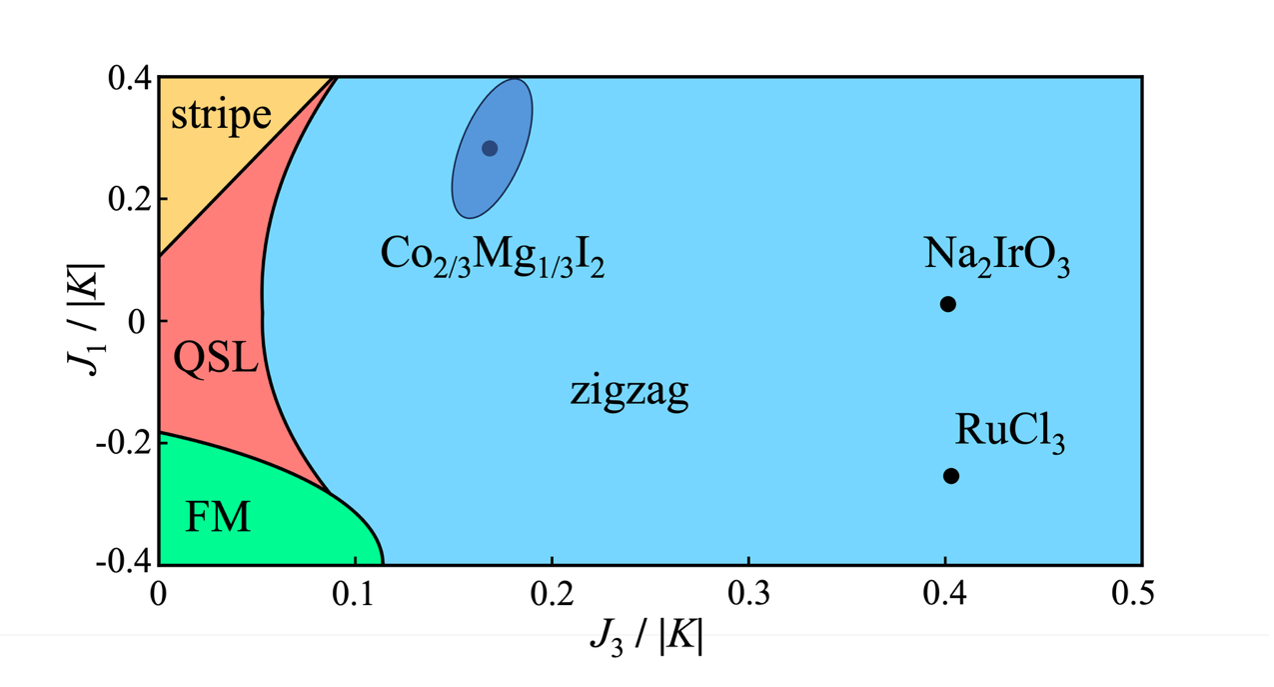}
	\centering
	\caption{Sensitivity of the Co$_{2/3}$Mg$_{1/3}$I$_2$ phase region to parameter variations. The dark blue area indicates the confidence interval for the Co$_{2/3}$Mg$_{1/3}$I$_2$ phase.
	}
  \label{hyperparameter}
  \end{figure}

   To assess the robustness of our conclusions, we performed a sensitivity analysis on Wannier-based extraction of crystal fields, $\Delta$ and $\delta$, and the selection of $\lambda$, $U$ and $J_\mathrm{H}$, by changing each parameters by 10\% as illustrated in \ref{hyperparameter}. We find that the variation does not affect our conclusion.

\begin{figure}
	\includegraphics[width=8cm]{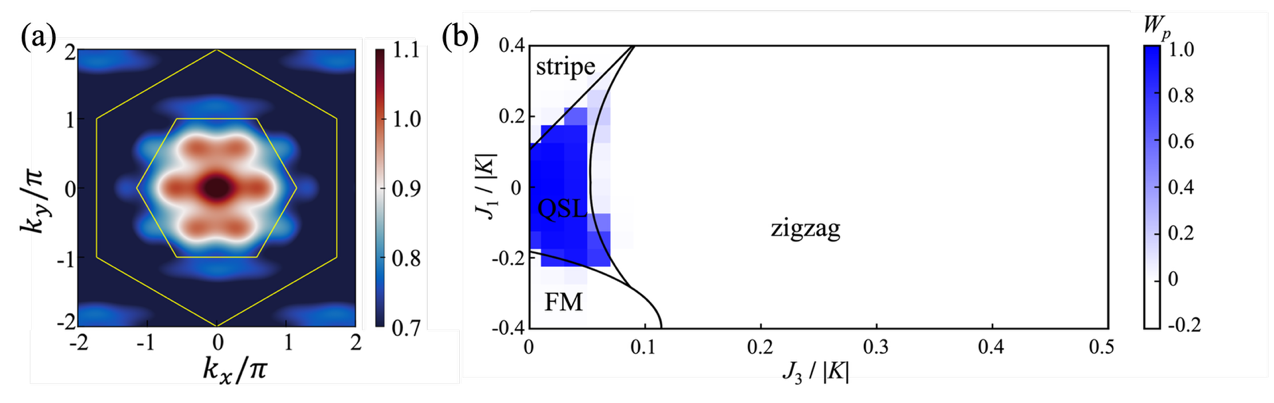}
	\centering
	\caption{(a) Magnetic structure factor of QSL region and (b) Phase diagram, obtained via exact diagonalization. The color scale indicates the expectation value of (a) magnetic structure factor, and (b) the plaquette operator $\langle W_p \rangle$.
	}
  \label{plaquette}
  \end{figure}

To determine the Kitaev QSL phase in Fig. \ref{hyperparameter}, we calculated the magnetic structure factor $S(q)$ of pure Kitaev model, as shown in \ref{plaquette}(a), and plaquette operator $\langle W_p \rangle$ in the phase diagram, as shown in \ref{plaquette}(b), via exact diagonalization method. The plaquette operator $\langle W_p \rangle$ is defined as the product of spin components around a hexagonal plaquette $W_p=2^6S_1^xS_2^yS_3^zS_4^xS_5^yS_6^z$, where the specific spin component at each site corresponds to the bond direction external to the loop, to character the Kitaev QSL\cite{Kitaev2006}. The lack of distinct peaks in Fig. \ref{plaquette}(a) and the computed values of $\langle W_p \rangle$ in Fig. \ref{plaquette}(b)  confirm the Kitaev QSL phase.

\begin{figure}
	\includegraphics[width=8cm]{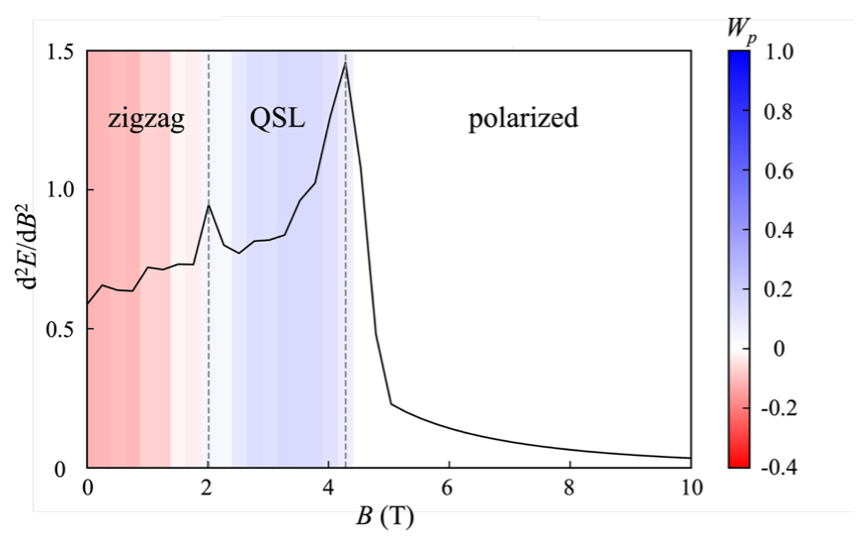}
	\centering
	\caption{Phase diagram of Co$_{2/3}$Mg$_{1/3}$I$_2$ obtained via exact diagonalization under external magnetic field. The color scale indicates the expectation value of the plaquette operator $\langle W_p \rangle$.
	}
  \label{magnetic_field}
  \end{figure}

To investigate the magnetic field response of Co$_{2/3}$Mg$_{1/3}$I$_2$, we incorporated the Zeeman term into the spin Hamiltonian as $H_\mathrm{Z}=g_\bot\mu_\mathrm{B}\mathbf{B}\mathbf{S}$, where $g_\bot$ denotes the out-of-plane $g$ factor and $\mu_\mathrm{B}$ is the Bohr magneton.  As seen in Fig. \ref{magnetic_field}, our calculations find that the Kitaev QSL phase emerges within the intermediate magnetic field of $2.0\mathrm{-}4.3\; \mathrm{T}$.c

\end{appendix}

\end{document}